\documentclass[journal=jacsat,manuscript=article,layout=singlecolumn]{achemso}
\usepackage{courier}
\usepackage[T1]{fontenc}
\usepackage[utf8]{inputenc}
\usepackage{lmodern}
\usepackage{amsmath} 
\usepackage{hyperref}
\usepackage{xcolor}
\usepackage{graphicx}
\usepackage{subcaption}
\usepackage{nomencl}

\newcommand{\EQ}{\begin{equation}}
\newcommand{\EN}{\end{equation}}
\newcommand{\EQA}{\begin{eqnarray}}
\newcommand{\ENA}{\end{eqnarray}}

\newcommand{\Eq}[1]{Eq.~(\ref{#1})}

\newcommand{\Seq}[1]{Section~\ref{#1}}
\newcommand{\Fig}[1]{figure~\ref{#1}}
\newcommand{\Tab}[1]{table~\ref{#1}}
\newcommand{\Tabs}[2]{tables~\ref{#1} and \ref{#2}}

\newcommand{\Da}{\rm Da}

\newcommand{\uu}{\boldsymbol u}
\newcommand{\FF}{\boldsymbol F}
\newcommand{\qq}{\boldsymbol q}
\newcommand{\SSSS}{\mbox{\boldmath ${\sf S}$} {}}

\newcommand{\Ox}{{\rm O}}

\newcommand{\CO}{{\rm CO}}
\newcommand{\COO}{{\rm CO_2}}
\newcommand{\CHHHH}{{\rm CH_4}}
\newcommand{\OO}{{\rm O_2}}
\newcommand{\NN}{{\rm N_2}}

\newcommand{\HHO}{{\rm H_2O}}
\newcommand{\HH}{{\rm H_2}}
\newcommand{\OH}{{\rm OH}}
\newcommand{\Hy}{{\rm H}}

\title{{High-resolution numerical simulations of turbulent non-catalytic reverse water gas shift}}

    \author{Nils Erland L. Haugen}
    \affiliation{SINTEF Energy Research, Trondheim, Norway}
    \alsoaffiliation{Division of Energy Science, Lule{\aa} University of Technology, Lule{\aa}, Sweden}
    \email{nils.e.haugen@sintef.no}
    \author{Axel Brandenburg}
    \affiliation{Nordita, Stockholm, Sweden}
    \author{Ewa Karchniwy}
    \affiliation{SINTEF Energy Research, Trondheim, Norway}
    \author{Ole Hauke Heinz Meyer}
    \affiliation{SINTEF Energy Research, Trondheim, Norway}
    \author{{\AA}smund Ervik}
    \affiliation{SINTEF Energy Research, Trondheim, Norway}
    \author{Hursanay Fyhn}
    \affiliation{SINTEF Energy Research, Trondheim, Norway}
    \author{Ladan Samaei}
    \affiliation{Nordic Electrofuel AS, Oslo, Norway}
    \author{Bj{\o}rn Bringedal}
    \affiliation{Nordic Electrofuel AS, Oslo, Norway}

    \SectionNumbersOn
\makenomenclature

\begin{document}
        \maketitle

\begin{abstract}
A green transition in aviation requires a drastic upscaling of Sustainable Aviation Fuel (SAF).
The power-to-liquid process for the production of CO$_2$-neutral jet fuel via electricity, called e-SAF, directly
replaces fossil jet fuel without having to change infrastructure, aeroplanes, or jet-engines.
The process combines green hydrogen with industrial exhaust gas, or captured carbon dioxide, in a circular economy concept.
A key element of the e-SAF production plant is the reactor where syngas is produced.
Traditional reactors use catalytic technology, which faces severe challenges
due to the reduced performance over time because of catalyst degradation, clogging, and breakup due to embrittlement.
This results in large operational costs and low process efficiency.
A high-potential alternative is the catalyst-free 
reverse water-gas-shift (RWGS) reactor concept, which has key
advantages in energy efficiency, flexibility of operation, potential
for upscaling, and applications in related fields.

The primary aim of this paper is to investigate the fundamental aspects
of the catalyst-free (non-catalytic) RWGS process, such as reaction kinetics and
the interactions between turbulence and chemistry. The secondary aim is to identify how a typical combustion subgrid scale models for Large Eddy Simulations (LES) perform when the chemical reactions are endothermic, in contrast to the strong endothermicity associated with classical combustion.
This is done by comparing results from LES with corresponding Direct Numerical Simulations (DNS),
which does not rely on any turbulence modeling.

It is found that even small traces of $\OO$ in the $\COO$ stream
can significantly increase the production rate of $\CO$.
This is attributed to the increased pool of $\OH$, which benefits the $\CO$ production rate.
The effect is strongest at atmospheric pressure and less pronounced at higher pressure.

By using the temporal jet framework to study turbulence-chemistry interactions,
an algebraic equation for the prediction of the $\CO$ conversion time in a
turbulent flow as a function of Damk\"{o}hler number and chemical timescale is employed.
This allows for an \textit{a priori} estimation of the required residence time to reach
full conversion of the RWGS reaction for various turbulent shear flows.

Finally, except for some smaller systematic deviations, it is concluded that the PaSR LES subgrid model designed for combustion reactions perform well also for the endothermic reverse water-gas-shift reaction.
\footnote{\today}
\end{abstract}

\newpage
	\tableofcontents
	\newpage
	\nomenclature{$\rho$}{Fluid density}
\nomenclature{$\uu$}{Fluid velocity}
\nomenclature{$P$}{Pressure}
\nomenclature{$\FF_{\rm visc}$}{Viscous force}
\nomenclature{$\SSSS$}{Traceless rate of strain tensor}
\nomenclature{$Y_k$}{Mass fraction of species $k$}
\nomenclature{$X_k$}{Mole fraction of species $k$}
\nomenclature{$J_k$}{Diffusive flux of species $k$}
\nomenclature{$\dot{\omega}_k$}{Reaction rate of species $k$}
\nomenclature{$T$}{Temperature}
\nomenclature{$R$}{Gas constant}
\nomenclature{$c_p$}{Heat capacity at constant pressure}
\nomenclature{$m_k$}{Molar mass of species $k$}
\nomenclature{$m$}{Mean molar mass of mixture}
\nomenclature{$h_k$}{Enthalpy of species $k$}
\nomenclature{$\qq$}{Heat flux}
\nomenclature{$\lambda$}{Thermal conductivity}
\nomenclature{$\kappa$}{PaSR correction factor for turbulence-chemistry interactions}
\nomenclature{$\tau_{\rm c}$}{Chemical timescale}
\nomenclature{$\tau_{\rm mix}$}{Timescale of turbulent mixing used in PaSR model}
\nomenclature{$\tau_{\rm f}$}{Fluid timescale, based on initial jet dimension and velocity}
\nomenclature{$C_{\rm mix}$}{Free parameter used to estimate $\tau_{\rm mix}$}
\nomenclature{Re}{Reynolds number}
\nomenclature{$z_{\rm H}$}{Mixture fraction based on hydrogen}
\nomenclature{$N_{\rm spec}$}{Number of species}
\nomenclature{FI}{Flame index}
\nomenclature{$[X]$}{Concentration of species X}
\nomenclature{$k$}{Reaction rate}
\nomenclature{$k_0$}{Pre-exponential factor for Arrhenius expression}
\nomenclature{$E$}{Activation energy}
\nomenclature{$\Da$}{Damkohler number}
\nomenclature{$V$}{Volume of grid cell}
\nomenclature{$V_{\rm domain}$}{Volume of computational domain}
\nomenclature{$L_j$}{Length of domain in direction $j$}
\nomenclature{$H_{\rm jet,u}$}{Initial height of jet based on velocity}
\nomenclature{$H_{\rm jet,sc}$}{Initial height of jet based on the scalars}
\nomenclature{$t_{\rm CO}$}{Time to reach 95\% yield of CO}
\nomenclature{$t_{\rm CO,app}$}{Time to reach 95\% yield of CO based on empirical expression}
\nomenclature{$\alpha$}{Empirical fitting value for expression for $t_{\rm CO,app}$}

	\printnomenclature[1.5cm]

	\section{Introduction}	

Carbon dioxide (CO$_2$), extracted directly from the atmosphere or
captured from industrial emissions, can be used as a source of carbon
for the production of useful organic chemicals, such as methane and other hydrocarbons.
Conversion of CO$_2$ into valuable products can be realized in different ways.
One way is to convert $\COO$ directly to methane through the Sabatier process:
    \begin{equation}
    \COO + 4\HH \Leftrightarrow \CHHHH + 2\HHO.
    \label{reac:methanation}
    \end{equation}    
A direct synthesis of methane through CO$_2$ hydrogenation is typically
characterized by a low CO$_2$ conversion and relatively low selectivity of
methane, even at high pressure and elevated temperature \cite{FAYISA2022}.
The requirement for a high $\HH$ concentration and operating pressure
makes the process expensive. 
Another approach involves converting CO$_2$ to CO by means of the reverse water–gas shift (RWGS) reaction, thereby producing synthesis gas (syngas), a mixture of CO and hydrogen:
    \begin{equation}
    \COO + \HH \Leftrightarrow \CO + \HHO
    \label{reac:rWGS}
    \end{equation}      
and further transform the products into synthetic fuels, such as methane and other hydrocarbons, for example via methanation and Fischer--Tropsch synthesis.
The RWGS reaction requires less $\HH$, and the produced CO gives a high flexibility for further transformation \cite{BENZINGER2019430,DESGAGNES2024152052}.
To achieve a high CO$_2$ conversion, a high temperature is required as
the RWGS reaction is endothermic, and thermodynamic equilibrium shifts
towards CO when the temperature increases.

The reaction is typically carried out in catalytic, fixed-bed reactors,
in which various catalysts can be used to enhance the reaction rate and CO selectivity.
In this context, a fixed-bed reactor means a chemical reactor that holds a stationary bed of solid particles through which gaseous reactants flow continuously.
In recent studies, a selectivity of nearly 100\% is demonstrated:
96.41\% by \citet{SANTOS2023118822} and 99.65\% by \citet{SUN201515985}.
However, catalytic reactors have to operate at moderate temperatures as
catalysts generally cannot withstand temperatures required to achieve
a high degree of CO$_2$ conversion due to the quick deactivation caused by
sintering \cite{Tang2021}.
Even with catalysts characterized by good thermal stability and with sophisticated reactor designs, under realistic conditions,
catalytic conversion usually does not exceed 50\%; 36.3\% was reported by \citet{SANTOS2023118822} and 49.7\% by \citet{SUN201515985}.
In addition, high residence times of reactants are required because of the relatively slow kinetics at moderate temperatures.
This usually means that multiple passes through the reactor are needed to achieve a satisfactory conversion degree. Furthermore, the catalytic activity of catalysts tend to decrease over time, making the process less efficient, and the preparation of catalysts with desired characteristics might be costly. 
In the past, most research efforts have focused on catalyzed RWGS reactions. Currently, the main interest within catalyzed RWGS are:
\begin{enumerate}
\item mathematical modeling and design of a catalytic reactor \cite{Ghodoosi2017,BENZINGER2019430,SANTOS2023118822},
\item development of inexpensive, synthetic catalysts with high selectivity,
stability and activity at low temperatures [examples of such novel catalysts are molybdenum phosphide \cite{Zhang2022},
cerium oxide \cite{Ebrahimi2022} or Ru clusters enclosed by silica shells \cite{Tang2021}], and
\item technological advances to intensify the process and shift the chemical equilibrium towards CO by, e.g., in situ H$_2$O removal \cite{DESGAGNES2024152052}, application of an external electric field \cite{Yamaoka2024}, or combining the RWGS reaction with chemical looping \cite{POMIRO2024112317}.
\end{enumerate}

The RWGS process can also be realized in a non-catalytic reactor fueled by a mixture of $\COO$ and $\HH$. In such reactors, the temperature is not limited by the stability of the catalyst. Therefore, the reactors can be operated at elevated temperature, which greatly increases the conversion rate and shortens the residence time.
The non-catalyzed alternative has not been much studied though.
In the past, Bustamante et al.\cite{Bustamante2004} investigated the process in quartz and Inconel\texttrademark\ 600 reactors
and formulated several global mechanisms for homogeneous RWGS reaction at different conditions. A novel non-catalyzed reactor concept was recently presented by Shekari et al. \cite{SHEKARI2023127947}, in which the RWGS reaction proceeds in a $\HH$ oxyflame, i.e., a flame resulting from feeding the reactor with streams of H$_2$ and O$_2$ + CO$_2$. At sufficiently high temperatures, conversion rates up to 79\% were reported for impressively low residence times of the order of 0.01\,s.

\section{Numerical modeling}
\subsection{DNS versus LES}
\subsubsection{Pencil Code -- the DNS solver}
The DNS in this work are performed using the \textsc{Pencil Code}, which is a multi-purpose, high-order, open-source code \cite{PC}.
The continuity equation is given by
\begin{equation}
\frac{D\rho}{Dt}=-\rho \boldsymbol{\nabla} \cdot \uu,    
\end{equation}
where $D/Dt=\partial/\partial t+\uu\cdot\boldsymbol{\nabla}$ is the advective derivative,
$\rho$ is the fluid density, and $\uu$ is the velocity vector.
The momentum equation reads
\begin{equation}
    \frac{D\uu}{Dt}=\frac{1}{\rho}
    \left(
    - \boldsymbol{\nabla} P + \boldsymbol{F}_{\rm vis}
    \right),
\end{equation}
where $P$ is the pressure and
\begin{equation}
    \boldsymbol{F}_{\rm vis}=\boldsymbol{\nabla} \cdot (2 \rho \nu \SSSS)
\end{equation}
is the viscous force with $\SSSS=\frac{1}{2}(\partial u_i/ \partial x_j + \partial u_j/ \partial x_i)
-\frac{1}{3}\delta_{ij}\boldsymbol{\nabla}\cdot\uu$ being the traceless rate-of-strain tensor
and $\delta_{ij} = 1\quad\text{if} \quad i=j \quad\text{and} \quad\delta_{ij} =  0 \quad\text{if} \quad i \neq j$ is the Kronecker delta.
The evolution of the mass fraction of species $k$ is given by
\begin{equation}
    \frac{DY_k}{Dt}=-\frac{1}{\rho}\left(
    \boldsymbol{\nabla}\cdot \boldsymbol{J}_k+\dot{\omega}_k
    \right),
\end{equation}
where $\boldsymbol{J}_k$ is the diffusive flux, which is calculated based on the mixture averaged approximation, and $\dot{\omega}_k$ is the reaction rate, both for species $k$.
Finally, the energy equation is solved with respect to the logarithm of the temperature $T$ as
\begin{equation}
    \frac{D\ln T}{Dt}=\frac{1}{c_p-R/m}\left[
    \sum_k \frac{DY_k}{Dt}\left(\frac{R}{m_k}-\frac{h_k}{T}\right)-\frac{R}{m}\boldsymbol{\nabla}\cdot\uu +\frac{2\nu\SSSS^2}{T}-\frac{\boldsymbol{\nabla}\cdot\qq}{\rho T}
    \right],
\end{equation}
where $c_p$ is the heat capacity at constant pressure, $R$ is the universal gas constant, $m_k$ is the molar mass of species $k$, $m$ is the mean molar mass of the mixture, 
\begin{equation}
    \qq=\sum_k h_k\uu-\lambda \boldsymbol{\nabla} T
\end{equation} 
is the heat flux, $h_k$ is the enthalpy of species $k$, and $\lambda$ is thermal conductivity.
Assuming an ideal gas, the pressure is found from
\begin{equation}
    P=\frac{\rho RT}{m}.
\end{equation}
For more details about the numerical implementation, the reader is referred to \citet{BABKOVSKAIA2011}.
The chemical kinetic mechanism used here is the one of \citet{LI2015617} with 14 species and 34 reactions. 

\subsubsection{OpenFoam -- the LES solver}
As for the \textsc{Pencil Code}, the LES solver
consists of a coupled set of partial differential equations describing the transient evolution of multicomponent turbulent reacting flow. To reduce computational cost, the LES model does not resolve the smallest wavelengths, which in this work are removed using Favre-filtered balance equations for mass, momentum, and energy.
The smallest scales are instead modeled using a dynamic one-equation subgrid-scale eddy viscosity model \cite{kim95}.
The eddy viscosity model dynamically computes the filter coefficients appearing in the LES formulation. 
We use the open-source C++ libraries OpenFOAM-12. OpenFOAM is a generic platform
for the finite-volume discretization of partial differential equations \cite{weller98}.
The compressible reactive \texttt{reactingFoam} solver is used.
Additionally, we employ the thermo-dynamically consistent \texttt{FickianTransportFoam} libraries provided by \citet{rintanen25}, appropriately accounting for non-unity Lewis number diffusivity effects.
The Lewis number is defined as the ratio of thermal to mass diffusivity.
This is particularly important for hydrogen-containing mixtures, such as those relevant for the RWGS, due to the large Lewis numbers of $\HH$ and $\Hy$. These large Lewis numbers result in strong mass diffusion of the same species. The importance of correctly describing, e.g., enthalpy diffusion for H$_2$ flames in shear layer turbulence is emphasized by \citet{cook09}. Additionally, the \texttt{FickianTransportFoam} model allows to specify inter-species diffusion coefficients for a range of pressures and temperatures, which are consistently combined to provide the species’ diffusion coefficient with respect to the mixture.

Turbulence-chemistry interaction is handled by the partially stirred reactor model (PaSR) \cite{fureby12}, assuming the chemical reactions to occur in topologically complex fine structures, making up a fraction of the volume in a given computational cell.
Reaction rates are scaled by a correction factor $\kappa$ to tune the turbulence-chemistry interaction.
In the PaSR formulation, this correction factor is estimated as
\begin{equation}
    \kappa = \frac{\tau_\mathrm{c}}{\tau_\mathrm{c} + \tau_\mathrm{mix}},
\end{equation}
where $\tau_\mathrm{c}$ approximates the chemical timescale and $\tau_\mathrm{mix}$ characterizes the turbulent mixing timescale.
The chemical timescale is calculated from the elementary reaction rates in the chemical mechanism,
while the turbulent mixing timescale is estimated based on characteristic turbulence quantities \cite{fureby12}.
For the OpenFOAM implementation of the PaSR model, $\tau_\mathrm{mix}$
depends linearly on the free parameter $C_\mathrm{mix}$, i.e., $\tau_\mathrm{mix} = C_\mathrm{mix} \sqrt{\nu_\mathrm{eff} / \epsilon}$, in terms of the effective turbulent viscosity, $\nu_\mathrm{eff}$, and the turbulence dissipation rate, $\epsilon$.
It can further be shown that
\begin{equation}
    C_\mathrm{mix} = \frac{1}{\sqrt{1 + 0.09 \,\mathrm{Re}}},
\end{equation}
in terms of the turbulent Reynolds number, $\mathrm{Re}$.
The coefficient $C_\mathrm{mix}$ can be used to modulate the burning rate predicted by the PaSR model.
Note that $C_\mathrm{mix} = 0$ gives $\kappa = 1$, i.e., full reactivity of each computational cell.
For the turbulence conditions at hand, $C_\mathrm{mix} = 0.01$ is used.
This is in line with previous work using the same computational setup for hydrogen combustion, e.g., \citet{heggset23b,heggset24,gruber24,meyer25}.
The low sensitivity to the $C_\mathrm{mix}$ parameter also for the present case is shown below (Sec.~\ref{sec:LES}).
The same detailed chemical reaction kinetics is used in the LES and the DNS.

In OpenFOAM, the PIMPLE algorithm, a hybrid SIMPLE-PISO iteration scheme, is used for pressure-velocity coupling,
allowing an adaptive time step that is limited by the user-specified Courant number $\sim 0.9$.
An implicit first order Euler scheme is used for time integration. Interpolation of the face fluxes to the cell values is achieved by combinations of second order central differencing schemes. 

To sum up, the DNS and the LES models in this work are largely distinguished through the resolution of the smallest turbulent scales,
the description of the turbulence-chemistry interaction, and the accuracy or order of the numerical schemes employed. 

\subsection{Some diagnostics measures}

The mixture fraction based on $\HH$ for the streams used in this work,
where the initial co-flow consist of only $\HH$ and $\HHO$ while there
are no hydrogen containing species in the initial jet, is defined as
\begin{equation}
    z_\Hy=\frac{Z_\Hy}{I_\Hy},
\end{equation}
where 
\begin{equation}
    I_\Hy=\frac{2m_\Hy}{m_{\HH}}Y_{{\rm init},\HH}+\frac{2m_\Hy}{m_{\HHO}}Y_{{\rm init},{\HHO}}
\end{equation}
and
\begin{equation}
    Z_H=\sum_{k=1}^{N_{\rm spec}} Y_k \frac{\mu_{H,k}m_\Hy}{m_k}.
\end{equation}
In the above, $N_{\rm spec}$ is the total number of species and $\mu_{H,k}$ is the number of hydrogen atoms in species $k$. In this way, $z_H$ is zero when there is no elemental hydrogen in the local gas, corresponding to the conditions of the initial central jet, and it is unity when the mixture equals the composition of the initial co-flow. 

Another useful quantity that can be utilized to better understand non-premixed flames is the flame index,
which is able to distinguish between reactions that proceed in premixed and non-premixed modes.
In particular, the flame index is negative when the gradients of the two
reacting species have opposite directions, i.e., they are reacting
before they have time to mix.
This corresponds to a large Damk\"{o}hler number.
On the other hand, for a propagating flame that reacts in a premixed mode, i.e., the Damk\"{o}hler number is small,
the reactants are mixed before the combustion reaction is initiated by the high temperature of the flame front.
In this situation, the gradients of the two reactants are in the same direction
(away from the temperature-induced flame front) and the flame index is therefore positive.
For the RWGS reaction, however, which, in contrast to the combustion reactions, is an endothermic reaction,
there will be no temperature-induced flame front.
Unlike for flames, the flame index will therefore be both positive and negative if the RWGS reaction proceeds under premixed conditions.
For large Damk\"{o}hler numbers, however, the flame index will be negative also for the RWGS reaction -- just as it is for flames.

The jet mixture in this study consists of both $\COO$ and $\OO$, which means that the flame index can be defined in two ways. 
The O$_2$-based flame index is defined as
\begin{equation}
    {\rm FI_{\OO}}=|\dot{\omega}_{\OO}|\frac{\boldsymbol{\nabla} Y_{\HH}\cdot \boldsymbol{\nabla} Y_{\OO}}{|\boldsymbol{\nabla} Y_{\HH}|\cdot |\boldsymbol{\nabla} Y_{\OO}|},
\end{equation}
while the one based on $\COO$ is given as
\begin{equation}
\label{eq:FICO2}
    {\rm FI_{\COO}}=|\dot{\omega}_{\COO}|\frac{\boldsymbol{\nabla} Y_{\HH}\cdot \boldsymbol{\nabla} Y_{\COO}}{|\boldsymbol{\nabla} Y_{\HH}|\cdot |\boldsymbol{\nabla} Y_{\COO}|}.
\end{equation}
Since the reactivity of $\HH$ with $\OO$ is very fast, this reaction will primarily proceed in a non-premixed mode.
Because of this, and since the main reaction in this study is the reaction between $\HH$ and $\COO$,
we will in the following only consider the $\COO$-based flame index as defined by \Eq{eq:FICO2}.

\subsection{Kinetic reaction mechanisms}	
\label{seq:kinetic}
    Since only catalytic reactors are commercially available and the research focuses primarily on catalyzed RWGS reaction, a vast majority of chemical mechanisms were developed for the catalyzed RWGS reaction in different conditions and for different catalysts \cite{QI2020131}. They differ in complexity as the catalytic effect can be described by simple power law expressions, Langmuir-Hinshelwood type kinetics or intrinsic kinetics \cite{LALINDE2020124629,Wolf2017}. Since the presence of catalysts affects the overall reaction rate, a different reaction mechanism is needed for each distinct catalyst, and for the same reason these mechanisms cannot be used to study uncatalyzed (homogeneous) RWGS reaction. 
    
    Several global mechanisms for uncatalyzed RWGS reaction were formulated in the past. A summary of those is given in \citet{Bustamante2004}. All the global expressions are in the form of:
    \begin{equation}
    \frac{d[\CO]}{dt} = k [\HH]^{1/2} [\COO],
    \label{eq:global}
    \end{equation}
    where $[\:]$ means concentration, $k = k_0 \exp{(-E/RT)}$ is the Arrhenius expression in which $k_0$ is the pre-exponential factor, $E$ is the activation energy, and $R$ and $T$ are the gas constant and temperature, respectively. The pre-exponential factor and activation energy are fitted to reproduce the experimental results, while the exponents for $\HH$ and $\COO$ concentrations are derived based on the Bradford mechanism \cite{Bradford1933}:
    \begin{align}
    \HH + M &\Leftrightarrow 2\Hy + M, \label{eq:Bradford1} \\ 
    \Hy + \COO &\Leftrightarrow \CO + \OH, \\
    \OH + \HH &\Leftrightarrow \HHO + \Hy
    \label{eq:Bradford3}
    \end{align}       
    under the assumption of steady state for H and OH, and equilibrium for the $\HH$ di/re-association. The Bradford mechanism itself can also be used to describe RWGS reaction although no kinetic parameters are given in \cite{Bradford1933} - the article in which the mechanism is introduced. A complete set of kinetic parameters for the Bradford mechanism was recommended by Kaskan \& Browne \cite{Kaskan1964} and their kinetic data for reactions \ref{eq:Bradford1}--\ref{eq:Bradford3} will in the following be referred to as the Bradford mechanism. It should be noted that the global expressions derived based on the Bradford mechanism have several limitations as they are strictly applicable only for low CO$_2$ conversion, and they are not consistent in a thermodynamical sense. Since in older experiments only a few percent of CO$_2$ conversion is reached, most often only coefficients for the forward reactions are provided. This means that the global expressions cannot predict a correct equilibrium composition, as will be shown in the next section.  

There seems to be no detailed mechanisms developed specifically for the homogeneous RWGS reaction. However, many syngas combustion mechanisms are available, e.g. \cite{KEROMNES2013995,DAVIS20051283, FRASSOLDATI20073471,LI2015617,LI2022112093,STARIK2010,SUN2007439}, and these should be suitable for modeling of CO$_2$ conversion via the RWGS reaction.
    
In order to gain confidence that the syngas mechanisms are appropriate for our purpose, their predictions should be compared with relevant experiments or, in the absence of complete dataset required for direct comparison (which is the case here), a comparison with global mechanisms formulated based on experimental results might be a viable alternative. First, we attempt to validate mechanisms by performing zero-dimensional simulations in Cantera \cite{Cantera} and comparing predictions of several syngas and global mechanisms. An attempt to reproduce experimental results of Graven \& Long \cite{Graven1954} is also made. Next, implementation of the selected mechanisms into the \textsc{Pencil Code} is verified by comparison with Cantera tests. SI units are used for all results presented in this paper.
    
Four syngas combustion mechanisms of similar complexity are selected for zero dimensional tests: Keromnes et al.~\cite{KEROMNES2013995}, Davis et al.~\cite{DAVIS20051283}, Frassoldati et al.~\cite{FRASSOLDATI20073471} and Li et al.~\cite{LI2015617}. An isothermal and isobaric \href{https://cantera.org/dev/python/zerodim.html}{\color{blue}{reactor}} available in Cantera was used for kinetics simulation with the selected mechanisms. The predictions were compared with five one-step global mechanisms specific for the RWGS reaction. The global mechanisms were formulated based on a few different sets of experimental results and are summarized by Bustamante et al. \cite{Bustamante2004}. 
    
\begin{figure}[h!]
        \centering
        \includegraphics[trim={0 8.5cm 0 1.5cm},clip,width=1.0\textwidth]{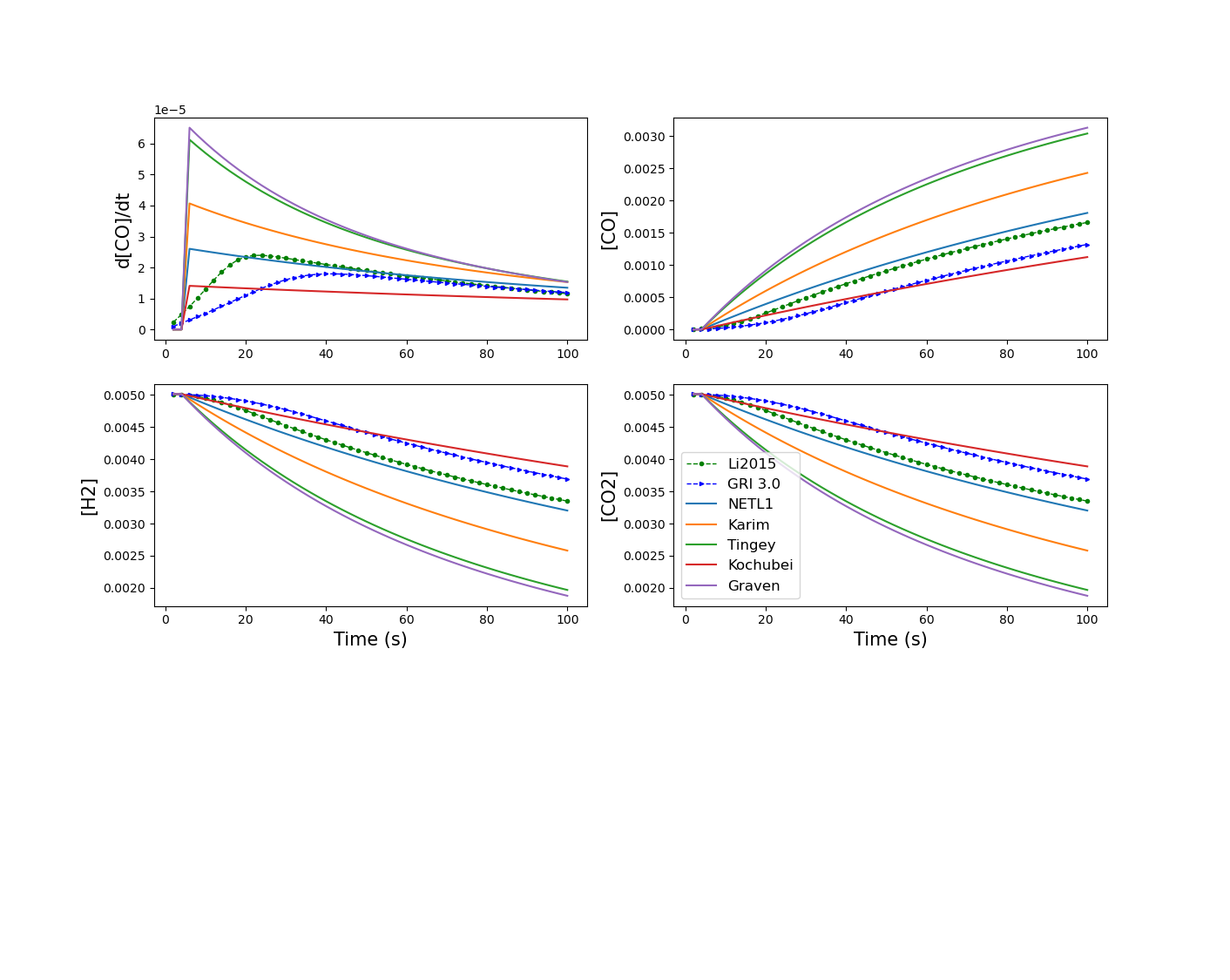}
        \caption{Comparison of species concentrations (kmol/m$^3$) for global and selected detailed mechanisms in conditions corresponding to experiment of Bustamante et al. \cite{Bustamante2004}. The two detailed mechanisms are marked with green and blue symbols.}
        \label{fig:test_mechanisms}
    \end{figure}

\begin{figure}[h!]
        \centering
        \includegraphics[trim={0 9cm 0 0.5cm},clip,width=1.0\textwidth]{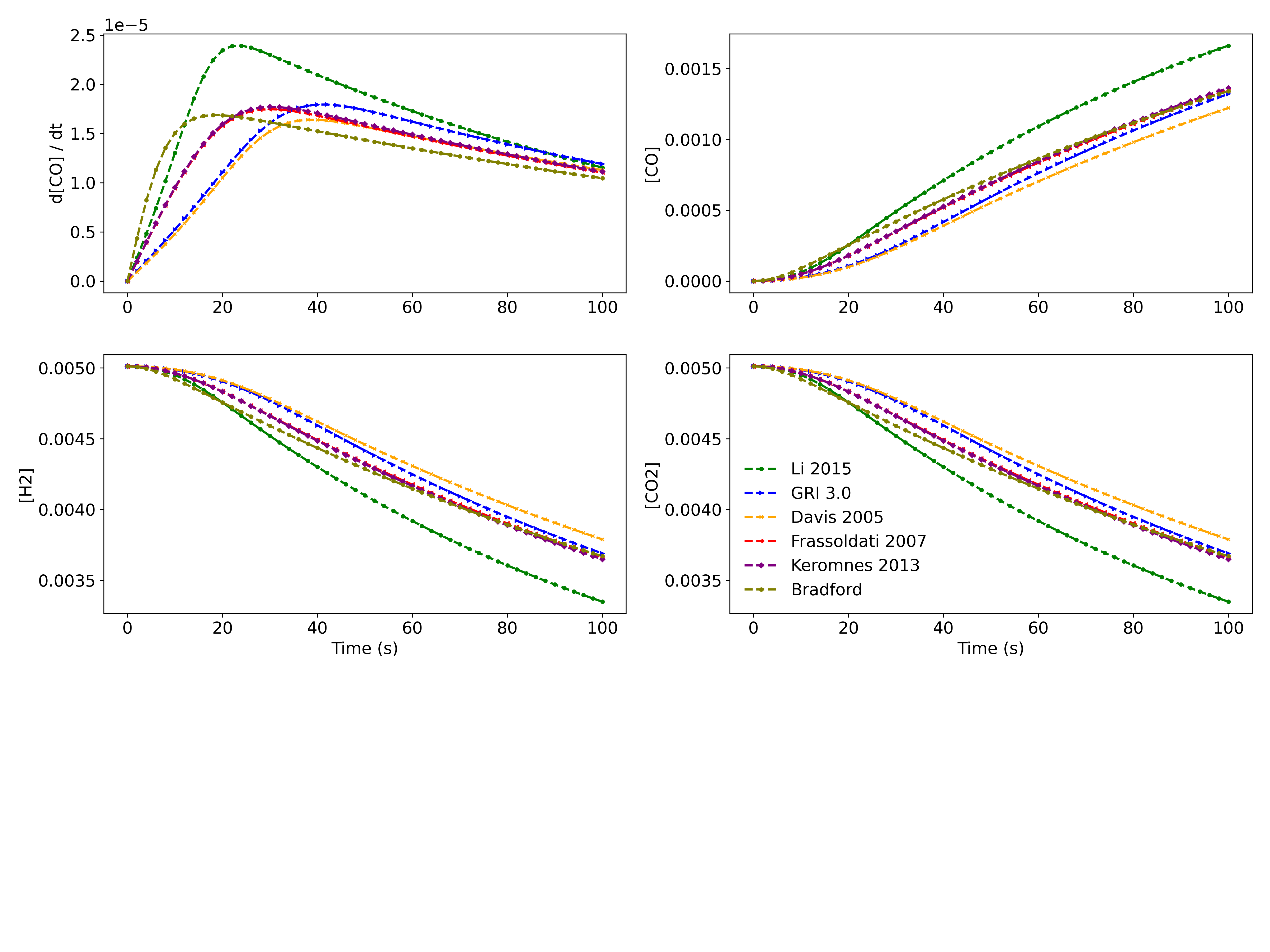}
        \caption{Comparison of syngas mechanisms predictions of species concentrations (kmol/m$^3$) in conditions corresponding to experiment of Bustamante et al. \cite{Bustamante2004}}        
        \label{fig:detailed_mechanisms}        
    \end{figure}

For reference, results obtained using a more complex, widely-used mechanism for methane combustion, GRI 3.0, are also included in the study. Fig~\ref{fig:test_mechanisms} presents a comparison of global mechanisms and two selected detailed syngas mechanisms, while in Fig.~\ref{fig:detailed_mechanisms} predictions of the other syngas mechanisms and GRI 3.0 are compared for the same conditions. As can be seen in Fig.~\ref{fig:test_mechanisms}, the discrepancy between global mechanism predictions is rather significant, they are therefore not a reliable source of validation data. Tingey \cite{Tingey1966} associates this discrepancy with small, immeasurable amounts of $\OO$ that can have a great impact on the reaction rate - this effect will be further investigated later in this document. The detailed mechanisms are in the best agreement with the mechanisms formulated by Tingey \cite{Tingey1966}, at least for the studied conditions, which correspond to the conditions
in which the experiment of Bustamante et al.\cite{Bustamante2004} was carried out ($T = 1200\,$K, $P = 1\,$bar and $\COO/\HH = 0.5/0.5$ (vol.)).
Since the detailed mechanisms account for radicals formation, the predicted formation of CO/H$_2$O and consumption of H$_2$/CO$_2$ is initially slower than what the global expressions predict. The mechanisms of Keromnes et al. \cite{KEROMNES2013995}, Davis et al. \cite{DAVIS20051283} and Frassoldati et al. \cite{FRASSOLDATI20073471} give very similar results to GRI 3.0, while the mechanism of Li et al. \cite{LI2015617} predicts slightly faster CO$_2$ conversion for the initial 50 seconds, as shown in Fig.~\ref{fig:detailed_mechanisms}. For the studied conditions, the Bradford mechanism \cite{Bradford1933} is in a very good agreement with detailed syngas mechanisms.


\begin{figure}[h!]
        \centering
        \includegraphics[trim={0 9cm 0 0.5cm},clip,width=1.0\textwidth]{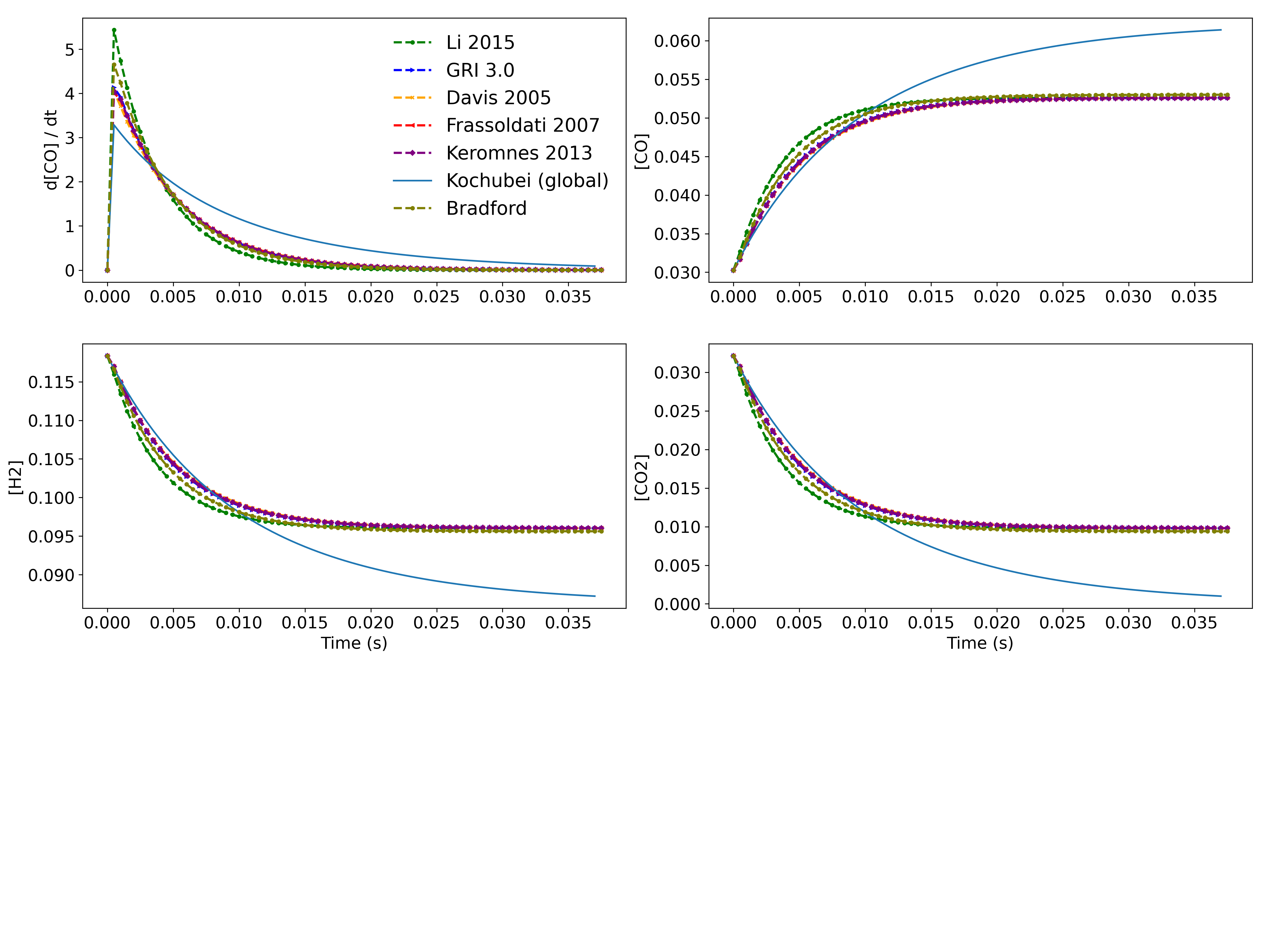}
        \caption{Comparison of selected mechanisms predictions of species concentrations (kmol/m$^3$) in conditions relevant for the studied reactor}        
        \label{fig:reactor_conditions} 
    \end{figure}

Fig.~\ref{fig:reactor_conditions} shows the same results obtained for the conditions listed in Table~\ref{tab:conditions}, which corresponds to typical industrial operating conditions. In these conditions, the conversion is several orders of magnitude faster due to higher temperature and pressure. All detailed mechanisms, and the Bradford mechanism, give, again, very similar results and converge towards the same equilibrium composition. The global expression is applicable only for the initial 0.01 seconds, after which it deviates significantly from the predictions of the detailed mechanisms due to the fact that the reaction progresses until all CO$_2$ is consumed, i.e. the global expression without the reverse reaction rate cannot predict the equilibrium state.

    \begin{table}[h!]
    \centering
    \begin{tabular}{cccccccc}
     T [K] & P [Pa] & X$_\HH$ & X$_\HHO$ & X$_\COO$ & X$_\CO$ & X$_\NN$ & X$_\OO$ \\   
     \hline
     1650 & 3e6 & 0.5415 & 0.16 & 0.147 & 0.1385 & 0.0135 & 0.0
    \end{tabular}
    \caption{\label{tab:conditions}Conditions relevant for the considered reactor}      
    \end{table}

\begin{figure}[h!]
        \centering
        \includegraphics[width=0.8\textwidth]{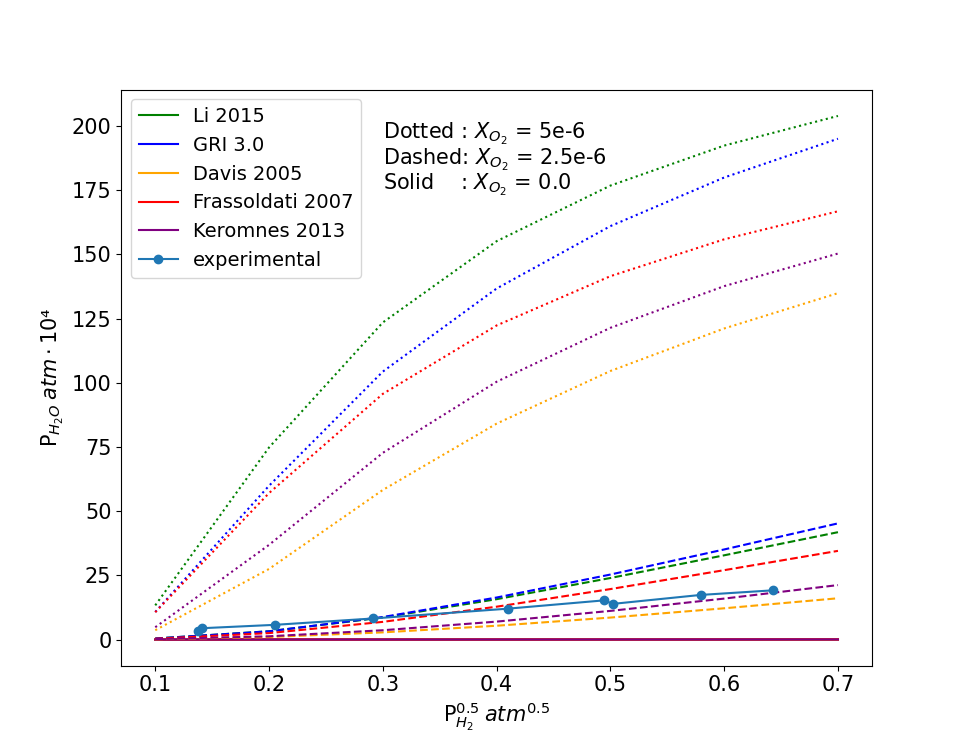}        
        \caption{Predictions of syngas mechanisms for the conditions corresponding to the experiment of Graven \& Long \cite{Graven1954}}
        \label{fig:validate_Graven}
    \end{figure}

In an attempt to validate the syngas mechanisms against experimental data, a case corresponding to the experiment performed by Graven \& Long \cite{Graven1954} was set up. In the experiment a stream of $\HH$, $\COO$ and $\NN$ was introduced into a cylindrical quartz reactor. The CO$_2$ partial pressure was kept constant, while the H$_2$ flow rate was varied. The total pressure was kept constant by adjusting the $\NN$ flow rate. The temperature inside the reactor was kept constant at 1173 K and the composition was measured at the reactor outlet after a flow time of 0.53 s. The results are presented in Fig.~\ref{fig:validate_Graven} in the form of the resulting H$_2$O partial pressure (as computed after 0.53 s) as a function of inlet H$_2$ partial pressure. The authors of the experiment claim that the reacting mixture was completely purified from $\OO$. If this is truly the case, the results obtained using all tested syngas mechanisms differ by 3 orders of magnitude from the experimental data (solid lines in Fig.~\ref{fig:validate_Graven} corresponding to X$_{\OO}$ = 0.0). However, if the measurements of the $\OO$ fraction were less accurate than six significant digits, then it is very likely that trace amounts of $\OO$ were present in the reactor. These small amounts of O$_2$ could have a significant impact on the results, as can be seen in Fig.~\ref{fig:validate_Graven} for the cases with X$_{\OO}$ = 2.5$\cdot 10^{-6}$ (dashed lines) and X$_{\OO}$ = 5$\cdot 10^{-6}$ (dotted lines). In fact, Graven \& Long claimed that 1\% of $\OO$ in the mixture accelerated the reaction 100-fold -- it is indeed predicted by all syngas mechanisms that the reaction rate becomes 20--800 times faster (depending on the initial H$_2$ partial pressure and mechanism) when the $\OO$ fraction increases from 2.5$\cdot 10^{-6}$ to 1$\cdot 10^{-2}$. It should be noted that the mechanism of Bradford~\cite{Bradford1933} was excluded from considerations in Fig.~\ref{fig:validate_Graven} as it does not include $\OO$, and therefore cannot account for the 
effect of $\OO$ on the reaction rate. 

By analyzing the net progress rates of reactions and species fractions at the end of simulation for the cases in which X$_{\OO}$ = 0 and X$_{\OO}$ = 2.5$\cdot 10^{-6}$, the strong effect that small amounts of $\OO$ has on the reaction will be explained in the following. 

The presence of $\OO$ causes much quicker formation of the OH radical, especially through the reaction 
\begin{equation}
\Hy + \OO \Rightarrow \Ox + \OH.
\label{reac1}
\end{equation}
This reaction does not happen when X$_{\OO}$ = 0 but it is among the fastest progressing reactions when X$_{\OO}$ = 2.5$\cdot 10^{-6}$. Additionally, substantial amounts of OH are produced through reactions of H, H$_2$, H$_2$O, HO$_2$ with atomic oxygen radical ($\Ox$), produced from O$_2$ dissociation. As a result, in the presence of $\OO$, the concentration of the OH radical is much higher - about 3-4 orders of magnitude in the considered case - than for the case with X$_{\OO}$=0. Subsequently, the fastest of all reactions, 
\begin{equation}
\HH + \OH \Rightarrow \Hy + \HHO,
\label{reac2}
\end{equation}
progresses a few orders of magnitude faster when $\OO$ is present in the mixture. The atomic hydrogen radical ($\Hy$) is then consumed according to the reaction
\begin{equation}
\COO + \Hy \Rightarrow \CO + \OH,
\label{reac3}
\end{equation}
the progress rate of which is almost equal to the progress rate of reaction \ref{reac2}. Apart from $\CO$, more $\OH$ radical is produced through reaction \ref{reac3}, which in turn accelerates reaction \ref{reac2}. Based on the potentially strong effect that even very small amounts of $\OO$ have on the reaction rate, it is tempting to call it a catalytic effect, but this is not appropriate since the $\OO$ is actually consumed in the process. Instead, we call it the oxygen derived free-radical promoted effect.

The free-radical promoted effect of $\OO$ makes validation of the selected mechanisms very challenging. Only experiments with extremely accurate measurements of $\OO$ fraction provide useful data for validation and the accuracy of such measurements is rarely stated. As we are unaware of experimental results with the required accuracy for $\OO$ concentration measurements, it seems reasonable to trust that the well-known and widely-used mechanisms optimized for syngas combustion are also suitable for modelling of the RWGS reaction. Since the predictions of all detailed mechanisms are similar, and given that the stiffness of the mechanisms is also similar, it is reasonable to choose the one with the lowest number of species and reactions as it will minimize the computational cost of of the resolved simulations. The number of species and reactions for the selected syngas mechanisms, as well as the time step required due to mechanism stiffness, are shown in Table~\ref{tab:SpeciesReactions}, from which it can be concluded that it will be computationally least expensive to employ the mechanisms of Li et al.\cite{LI2015617} or Davis et al.\cite{DAVIS20051283}. However, if the free-radical promoted effect of $\OO$ is insignificant, and a real reaction pathway is not of interest, it will by far be cheapest to use the mechanism of Bradford \cite{Bradford1933,Kaskan1964}. For the rest of this paper, unless otherwise specified, we will use the mechanism of Li et al.
    \begin{table}[h!]
    \centering
    \begin{tabular}{cccc}
      Mechanism & Number of species & Number or reactions & Required time step\\   
     \hline
      Li et al. \cite{LI2015617} & 14& 37 & 2e-9\\   
      Davis et al. \cite{DAVIS20051283} & 14&38 & 2e-9 \\   
      Keromnes et al. \cite{KEROMNES2013995} & 15& 48 & ND\\   
      Frassoldati et al. \cite{FRASSOLDATI20073471} & 21& 62 & ND\\         
      GRI 3.0 &53 & 325 & 2e-9\\   
      Bradford \cite{Bradford1933,Kaskan1964} &6 & 3 & 1e-9\\         
    \end{tabular}
    \caption{Number of species and reactions for the selected mechanisms}  
    \label{tab:SpeciesReactions}  
    \end{table}

\section{Results}
\subsection{Premixed zero-dimensional simulations}

 \begin{figure}[b!]
        \centering
        \includegraphics[width=1.0\textwidth]{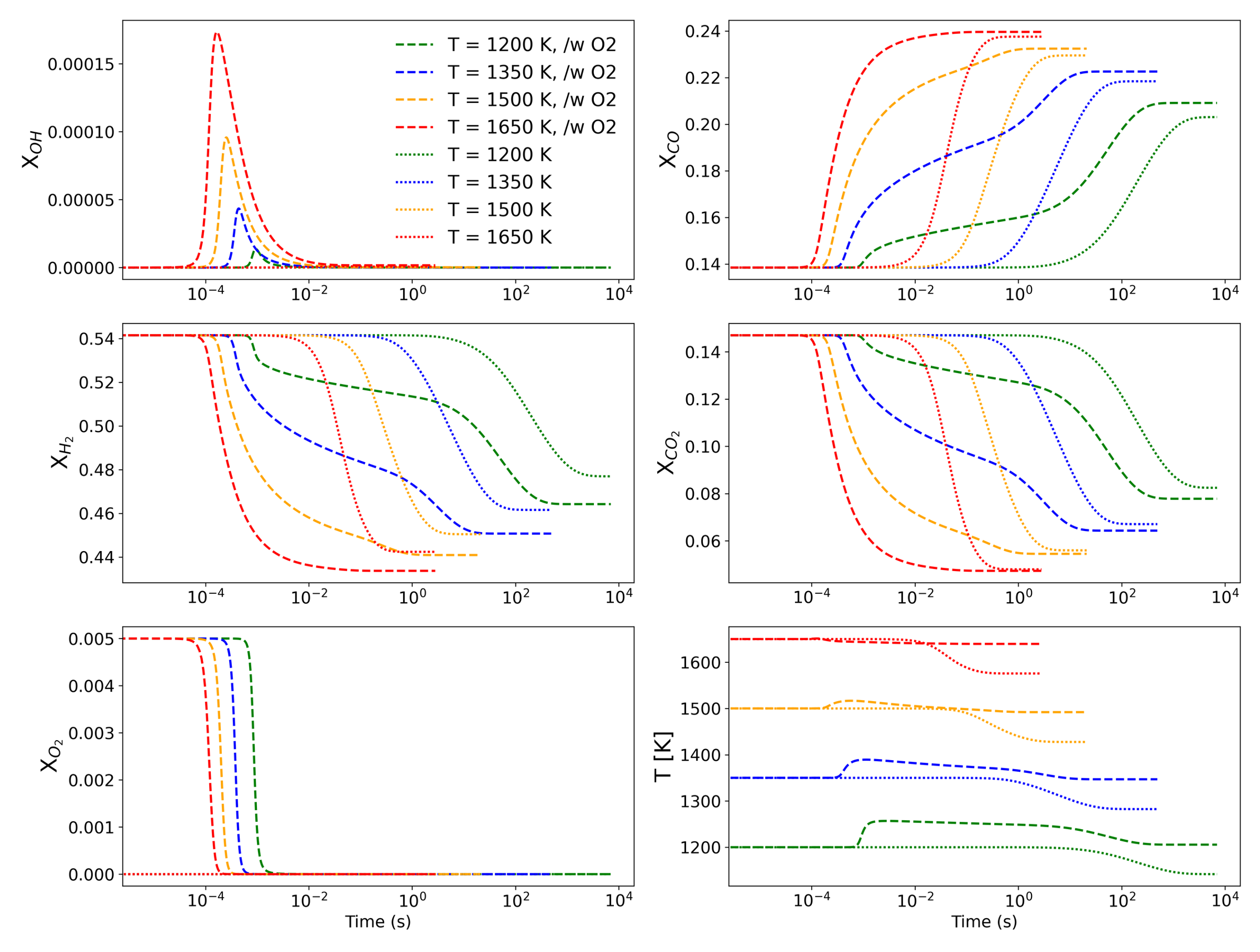}
        \caption{Effect of 0.5\% volumetric O$_2$ content at atmospheric pressure.}
        \label{fig:cat_1atm}
    \end{figure}   

     \begin{figure}[h!]
        \centering
        \includegraphics[width=1.0\textwidth]{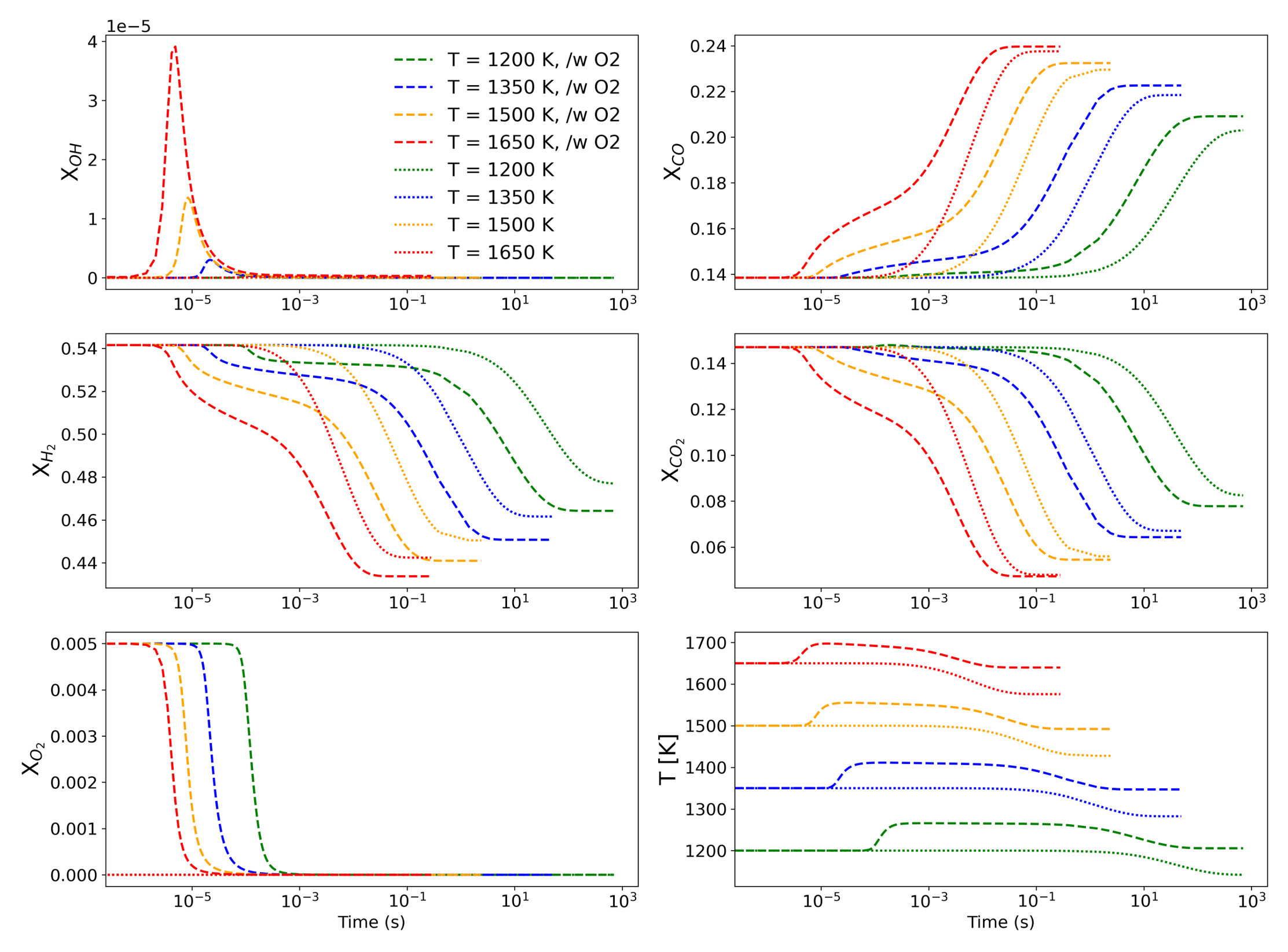}
        \caption{Effect of 0.5\% volumetric O$_2$ content at 30 bar.}
        \label{fig:cat_30atm}
    \end{figure}    

The increased rate of the RWGS reaction due to the presence of trace amount $\OO$ might be very beneficial for the reactor performance as much faster conversion means much shorter residence times, which can reduce the overall size of the reactor. In order to explore potential benefits of the free-radical promoted effect of $\OO$, additional zero-dimensional simulations are performed in Cantera. This time, only one mechanism, the mechanism of Davis et al. \cite{DAVIS20051283}, is used, and the simulations are performed for temperatures that are likely to occur in the reactor,
ranging from 1200 to 1650\,K.
The composition is the same as given in Table~\ref{tab:conditions}, with a difference that in half of the cases 0.5\% of N$_2$ is replaced by O$_2$. Simulations are performed both for the atmospheric pressure and for p = 30 bar, to get a sense how important the effect of $\OO$ is in conditions of interest (the reactor is planned to be operated at high pressure) relative to the atmospheric pressure, at which the experiment of Graven \& Long \cite{Graven1954} was performed. Fig. \ref{fig:cat_1atm} presents the results for atmospheric pressure, while Fig. \ref{fig:cat_30atm} presents the same results for high pressure. In the figures the cases with no O$_2$ are marked by dotted lines, while the cases with 0.5\% volumetric O$_2$ content are marked by dashed lines. As can be seen in Fig. \ref{fig:cat_1atm}, the effect of added $\OO$ is strongest for highest temperature, $T = 1650\,$K, at which the equilibrium state is reached around two orders of magnitude faster in the presence of $\OO$, compared to the cases without O$_2$. Despite the fact that for lower temperatures conversion for the cases with $\OO$ begins about three orders of magnitude faster than without $\OO$, the speed-up in the total conversion, at which the equilibrium state is reached, is not as large. The most likely reason for that is that for higher temperatures much more OH radicals is produced (see upper left panel of Fig. \ref{fig:cat_1atm}), which sustains faster conversion rate for the entire conversion duration. For cases with lower temperatures, the conversion rate decreases when OH radicals are consumed after around 0.001 - 0.01 s, which can be seen as flattening of the conversion curves for CO, H$_2$ and CO$_2$. Looking now at the corresponding results for the cases with higher pressure, Fig. \ref{fig:cat_30atm}, it is clear that the free-radical promoted effect of $\OO$ is weaker than at atmospheric pressure. At this pressure, however, the effect of O$_2$ is much less dependent on temperature, and the effect is strongest for lowest temperature, at which the equilibrium state is reached about one order of magnitude earlier in the presence of $\OO$. This change is probably due to significantly smaller amount of OH radicals that are created at high pressure. In conclusion, for the range of temperatures that are likely to be present in the considered reactor, the free-radical promoted effect of $\OO$ in high-pressure conditions is expected to be weaker than at atmospheric pressure. Nevertheless, the effect is still significant and has the potential to benefit the reactor performance.

\subsection{Temporal jet simulations}
\begin{figure}[h!]
        \centering
        \includegraphics[width=0.8\textwidth]{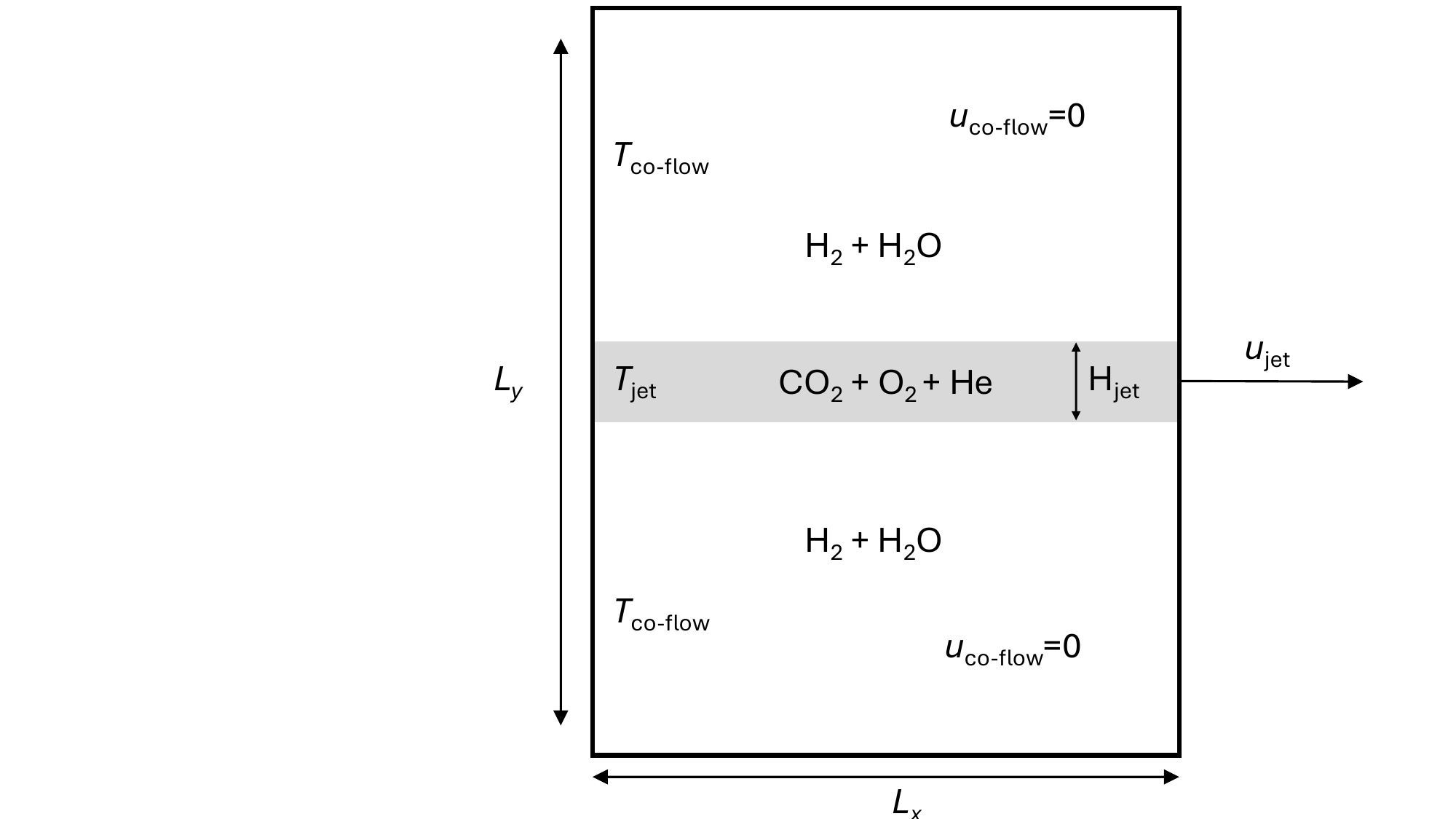}
        \caption{Sketch of initial condition for the temporal jet setup.}
        \label{fig:scetch_temporal_jet}
    \end{figure}

The aim of this work is to investigate the coupling between mixing and the endothermic RWGS reaction.
For simplicity and computational efficiency, the temporal jet framework \cite{HAWKES2007}, as presented in \Fig{fig:scetch_temporal_jet},
will be utilized to study the effect of turbulence on the RWGS reaction.
The initial conditions are homogeneous in the third direction, which has an extent of $L_z$.
All boundary conditions are periodic.

It should be noted that the velocity jet ($H_{{\rm jet},u}$) is slightly thinner than the scalar (temperature and species) jet ($H_{{\rm jet,sc}}$).
This has been done to avoid numerical problems due to the strong difference in viscosity and density between H$_2$ and CO$_2$.
Another modification that has been done to alleviate this problem is to add some helium to the CO$_2$ stream.
The helium does not have any impact on the reactions, but it increases the viscosity and lowers the density of the jet to make it more similar to that of the H$_2$-dominated co-flow. All simulations are performed for a pressure of 30 bar.
To study the effect of turbulence on the RWGS reaction, the Damk\"{o}hler number, given by the ratio of the fluid and chemical timescales, must be significantly smaller than unity. The Damk\"{o}hler number is defined as 
\EQ
\label{eq:Da}
\Da = \frac{\tau_{f}}{\tau_c},
\EN
where $\tau_f=H_{{\rm jet},u}/u_{\rm jet}$ is the timescale of the fluid based on the thickness and velocity of the jet, and $\tau_c$ is the chemical timescale, which is defined as the time it takes for a perfectly pre-mixed case to reach a $\CO$-yield of 95\% of the $\CO$-yield in equilibrium.
To make the DNS simulations computationally feasible, the computational domain (and the Reynolds number) has to be reduced in size compared to that of an industrial unit. 
The initial temperature of the simulation has therefore been increased to speed up conversion and make simulations feasible, and at the same time ensure that the Damk\"{o}hler number remain small enough such that the effect of turbulence on the RWGS reaction can be studied.
Somewhat arbitrarily, an initial temperature of 2000\,K has therefore been chosen for the benchmark simulations. 
In order to initiate turbulence in the shear layers, some randomness is required.
Here, randomness is introduced by adding low-amplitude initial Gaussian-distributed white noise in the velocity field.
The amplitude of the noise is homogeneous over the volume and is set to slightly below 1\% of the jet velocity.

\subsubsection{DNS}
Given appropriate boundary conditions and accurate chemical kinetics,
the DNS can be considered as numerical experiments because they resolve all temporal and spatial scales.
It is therefore very useful to use DNS as validation cases for Large Eddy Simulations (LES). In this section we will present three different three-dimensional simulations that are characterized by different Damk\"{o}hler and Reynolds numbers and that will later be used to validate the LES simulations. More specifications of these simulations are listed as Cases 3D1,  cis3D2 and 3D1-2600K in \Tabs{tab:params}{tab:params2}.
 \begin{table}[h!]
    \centering
    \begin{tabular}{l|ccc|cccc|c}
               & \multicolumn{3}{c}{Dimensions} &\multicolumn{4}{|c}{Jet} & \multicolumn{1}{|c}{Co-flow} \\
    Case       & $H_{{\rm jet},u}$ & $H_{\rm jet,sc}$ & $L_x, L_y, L_z$ & $T_{\rm jet}$ & $Y_{\OO}$ & $Y_{\NN}$ & $Y_{\COO}$ & $T_{\rm co-flow}$   \\ 
               &     [mm]          &        [mm]          &       [mm]  & [K]           &    -     &    -     &    -      &       [K]    \\  
     \hline
     \multicolumn{9}{c}{One-dimensional simulations} \\
     \hline
     1D1       & 0.134 & 0.2 & -, 2, - & 2000          &  0.054   &   0      &   0.746     &    2000     \\
     1D2       & 0.268 & 0.4 & -, 4, - & 2000          &  0.054   &   0      &   0.746     &    2000     \\  
     1D1noO2   & 0.134 & 0.2 & -, 2, - & 2000          &  0       &   0.054  &   0.746     &    2000     \\
     \hline
     \multicolumn{9}{c}{Two-dimensional simulations} \\
     \hline
     2D1       & 0.134 & 0.2 & 2, 2, - & 2000          &  0.054   &   0      &   0.746     &    2000     \\
     2D2       & 0.268 & 0.4 & 4, 4, - & 2000          &  0.054   &   0      &   0.746     &    2000     \\ 
     2D3       & 0.536 & 0.8 & 8, 8, - & 2000          &  0.054   &   0      &   0.746     &    2000     \\ 
     2D1noO2   & 0.134 & 0.2 & 2, 2, - & 2000          &  0       &   0.054  &   0.746     &    2000     \\
     2D1x2O2   & 0.134 & 0.2 & 2, 2, - & 2000          &  0       &   0.108  &   0.692     &    2000     \\
     2D1-1800K & 0.134 & 0.2 & 2, 2, - & 1800          &  0.054   &   0      &   0.746     &    1800     \\
     2D1-2200K & 0.134 & 0.2 & 2, 2, - & 2200          &  0.054   &   0      &   0.746     &    2200     \\
     2D1-2400K & 0.134 & 0.2 & 2, 2, - & 2400          &  0.054   &   0      &   0.746     &    2400     \\
     2D1-2600K & 0.134 & 0.2 & 2, 2, - & 2600          &  0.054   &   0      &   0.746     &    2600     \\
     \hline
     \multicolumn{9}{c}{Three-dimensional simulations} \\
     \hline
     3D1       & 0.134 & 0.2 & 2, 2, 1 & 2000          &  0.054   &   0      &   0.746     &    2000    \\
     3D2       & 0.268 & 0.4 & 4, 4, 2 & 2000          &  0.054   &   0      &   0.746     &    2000     \\
     3D1-2600K & 0.134 & 0.2 & 2, 2, 1 & 2600          &  0.054   &   0      &   0.746     &    2600     \\
    \end{tabular}
    \caption{\label{tab:params} Dimensions and initial parameters for the various simulations. In addition to what is listed in the table, all simulations have the following in common; the initial thickness of the shear layer is given by $\delta_{\rm shear}=0.02$ mm, which is smoothed with a hyperbolic profile between the jet and the co-flow, the mesh size, which is the same in all directions, is $\Delta x=4\times 10^{-3}$ mm, the velocity of the jet is $u_{\rm jet}=387.5$ m/s, the mass fraction of helium is 0.2 in the jet and zero in the co-flow, the mass fraction of steam is zero in the jet and 0.001 in the co-flow while the mass fraction of H$_2$ is 0 in the jet and 0.999 in the co-flow.}      
    \end{table}

 \begin{table}[h!]
    \centering
    \begin{tabular}{l|c|ccc|ccc}
                   &              & \multicolumn{3}{|c}{timescales}           & \multicolumn{3}{|c}{Non-dim. params}    \\
    Case           &max($Y_{\CO}$) & $\tau_f$      & $\tau_c$    & $t_{\CO}$     & Da     &      $t_{\CO}/\tau_c$ & Re      \\ 
                   &      -       & [$10^{-7}$s]  & [$10^{-7}$s]& [$10^{-7}$s] &  -     &             -        &  -      \\  
     \hline
     \multicolumn{8}{c}{Two-dimensional simulations} \\
     \hline
    2D1            & 0.294        &     3.46      &      590    & 797          & 0.0059 &        1.35          & 1875    \\
    2D2            & 0.294        &     6.92      &      590    & 830          & 0.0117 &        1.40          & 3750    \\
    2D3            & 0.294        &    13.83      &      590    &1100          & 0.0235 &        1.88          & 7500    \\
    2D1noO2        & 0.294        &     3.46      &     2880    & 3375         & 0.0012 &        1.17          & 1890    \\
    2D1x2O2        & 0.272        &     3.46      &      200    &  322         & 0.0173 &        1.61          & 1850    \\
    2D1-1800K      & 0.292        &     3.46      &     5510    &  617         & 0.0006 &       1.12           & 2240    \\
    2D1-2200K      & 0.295        &     3.46      &      105    &  279         & 0.033 &        2.66           & 1600    \\
    2D1-2400K      & 0.295        &     3.46      &       36    &  210         & 0.096 &        5.84           & 1390    \\
    2D1-2600K      & 0.295        &     3.46      &       18    &  206         & 0.198 &       11.79           & 1220    \\
     \hline
     \multicolumn{8}{c}{Three-dimensional simulations} \\
     \hline
     3D1            & 0.294        &     3.46      &      590    & NA          & 0.0059 &        NA            & 1875      \\
     3D2            & 0.294        &     6.92      &      590    & NA          & 0.0117 &        NA            & 3750      \\
     3D1-2600K      & 0.295        &     3.46      &       18    & 195         & 0.198 &       11.15           & 1220      \\
    \end{tabular}
    \caption{\label{tab:params2} Initial parameters for the various simulations. For all cases, the mass fraction of helium is 0.2 in the jet and zero in the co-flow. Furthermore, the mass fraction of $\COO$ is 0.746 in the jet and zero in the co-flow.}      
    \end{table}

\begin{figure}[h!]
        \centering
        \includegraphics[width=1.0\textwidth]{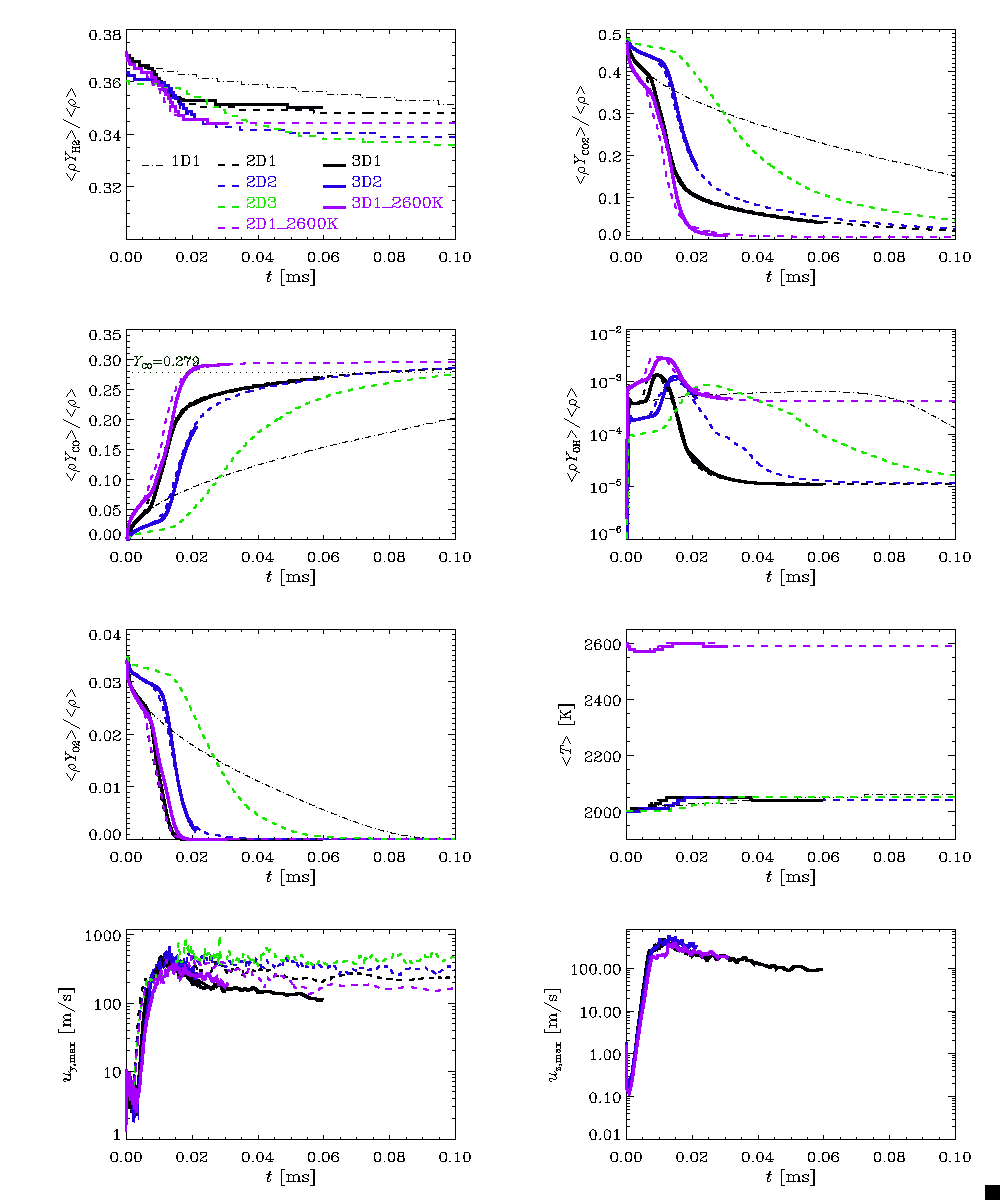}
        \caption{Results for some of the temporal jet simulations defined in \Tabs{tab:params}{tab:params2}.}
        \label{fig:comp_runs_2000Kb}
    \end{figure} 
In the following, the volumetric averaging of a variable $\phi$ is defined as
\begin{equation}
    \langle\phi\rangle=\frac{1}{V_{\rm domain}}\sum_{i,j,k}\phi(i,j,k)V,
\end{equation}
the summation is over all three-dimensional indexes $i,j,k$, $V$ is the volume of a grid cell and $V_{\rm domain}$ is the total volume of the domain.

In \Fig{fig:comp_runs_2000Kb}, we compare characteristic time traces of integral quantities obtained from the DNS. The benchmark three-dimensional simulation (3D1) is presented by the thick black solid lines. Inspecting the two lower panels of the figure, we see that it takes about 0.01ms for the perturbations in the flow to grow the transversal velocities to their maximum. After this initial period of velocity growth, the turbulent decay results in a slow decay in turbulence intensity. Comparing results from the three-dimensional results with the corresponding two-dimensional results (black dashed lines), we see that the growth periods of the two simulations are pretty similar, but the decay is much slower for the two-dimensional simulation. This is due to the fact that in contrast to the direct turbulent cascade to \emph{smaller} scales of a three-dimensional flow, a two-dimensional flow experience an indirect cascade to \emph{larger} scales. Since turbulent dissipation occurs at small scales, the dissipation of energy is therefore much slower for a two-dimensional flow. Comparing the benchmark case with a three-dimensional simulation with higher Reynolds number (3D2 - blue solid lines in the figure), we see that for the higher Reynolds number case the turbulent decay is somewhat slower than for the benchmark case -- as expected. The high-Reynolds number case is, however, too short to say anything very conclusive.

Moving now on to the mass fractions of $\OO$ and $\HH$, we see that they have a rather quick initial decrease, which becomes even steeper as soon as turbulence has developed. The initial decrease is due to the partially pre-mixed initial boundary layer (with thickness $\delta_{\rm shear}$) between the jet and the co-flow. As soon as the first turbulent eddies appear in the flow, the mixing due to them - and due to the following turbulence - quickly ensure that the $\OO$ in the jet mixes and reacts with $\HH$ in the co-flow, which results in a rather minor ($\sim 50$ K) increase in average temperature of the mixture. This oxidation of $\HH$ with $\OO$ generates a large pool of $\OH$, which in turn facilitates the conversion of $\COO$ to $\CO$, which is seen by the fact that the rate of $\CO$ production levels off at the time when the amount of $\OH$ in the mixture is reduced. This happens somewhat earlier for the  benchmark case than for the case with higher Reynolds number (3D2). This effect of $\OH$ on the $\CO$ production is presented in more detail in \Seq{seq:kinetic}. It is clear from these results that including some $\OO$ in the $\COO$ stream has a positive effect on the $\CO$ production rate.

\begin{figure}[h!]
        \centering
        \includegraphics[width=0.4\textwidth,trim={0cm 0cm 0 0cm},clip]{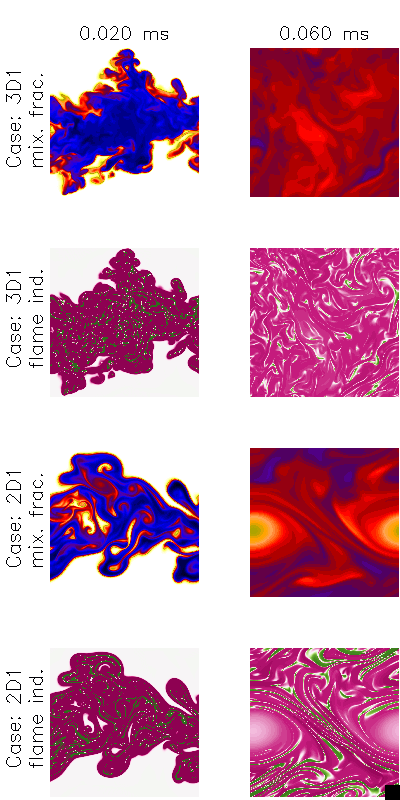}
        \caption{Cross sections of mixture fraction (rows 1 and 3) and flame index (rows 2 and 4) for early (left column) and late (right column) times. Results from three-dimensional simulations are shown in the two upper rows, while results from two-dimensional simulations are shown in the two last rows.}
        \label{fig:snapshots_6em5}
\end{figure}

It is interesting to note that the results from the three-dimensional simulations are remarkably similar to those of the corresponding two-dimensional simulations (dashed lines). From the contour plots of mixture fraction and flame index for corresponding two- and three-dimensional simulations visualized in \Fig{fig:snapshots_6em5}, we see that the similarity between the two simulations is relatively good at early times (0.02 ms) - before the turbulent cascade has become too important. At later times, however, the fundamental difference between the direct cascade of the three-dimensional flow and the inverse cascade of the two-dimensional flow becomes apparent by the scale difference of the dominating eddies. Later we will nevertheless utilize the early similarity between two and three dimensional simulations to perform affordable parameter studies using two-dimensional simulations. 

Reducing the dimensionality of the simulation further to one dimension (dashed-dotted lines), no turbulence or vortex shedding is possible. Hence, mixing of species is due only to molecular mixing. It is clear from the one-dimensional results (dashed-dotted lines) that the lack of turbulence results in a much slower conversion of CO$_2$ - as expected.

\begin{figure}[h!]
        \centering
        \includegraphics[width=1.0\textwidth,trim={0cm 0cm 0 0cm},clip]{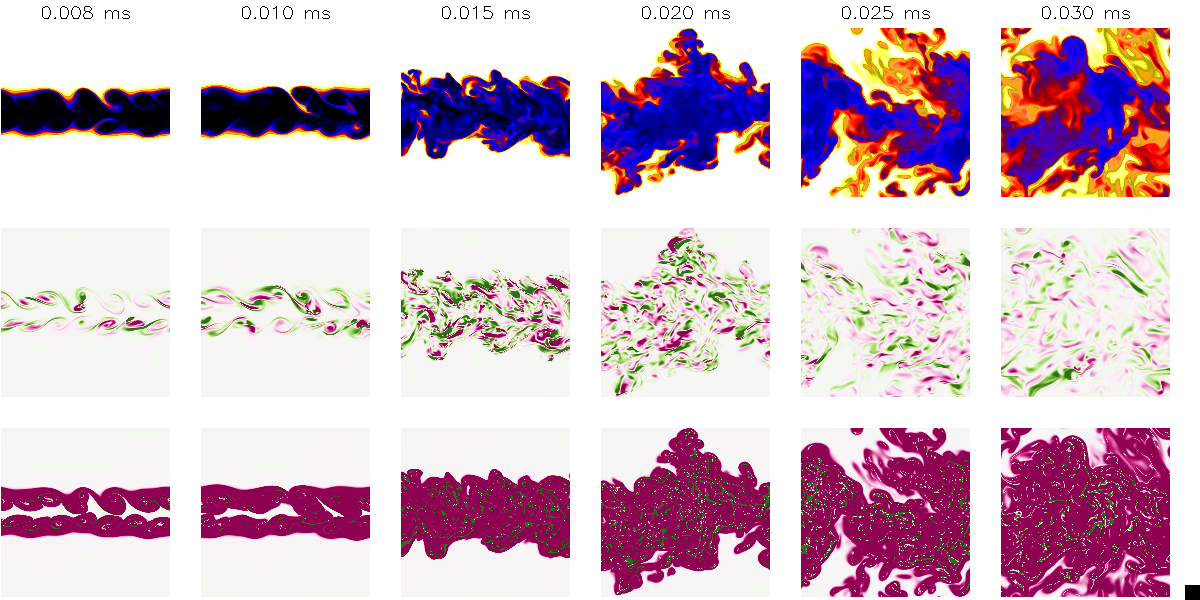}
        \caption{Cross section of mixture fraction (upper row), vorticity in $x$-direction (middle row) and flame index (lower row) for case 3D1. For the flame index, negative values are represented by purple color while positive values have green color.}
        \label{fig:snapshots_3D1}
\end{figure}

\begin{figure}[h!]
        \centering
        \includegraphics[width=1.0\textwidth,trim={0cm 0cm 0 0cm},clip]{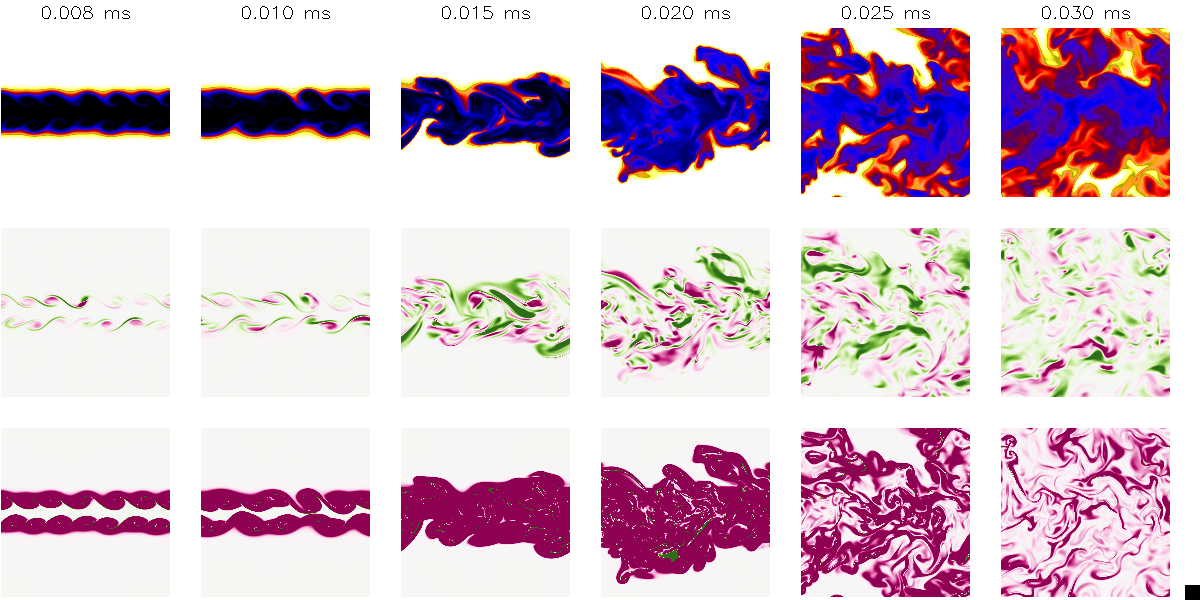}
        \caption{Cross section of mixture fraction (upper row), vorticity in $x$-direction (middle row) and flame index (lower row) for case 3D1-2600K. For the flame index, negative values are represented by purple color while positive values have green color.}
        \label{fig:snapshots_3D1_2600K}
\end{figure}

Contour plots of vorticity, mixture fraction and flame index are visualized in \Fig{fig:snapshots_3D1} for case 3D1. From the last panels it can be seen that the extent of the domain in the y-direction is too small to contain all the turbulent flow. Since all boundary conditions are periodic, this means that in reality these simulations correspond to the infinite row of jets in the y-direction, with the separation between each jet being equal to $L_y$. From the central panel, representing mixture fraction - with black corresponding to zero mixture fraction and white being a mixture fraction of unity - we see that after about 0.03 ms there is no pure $\HH$ left in the domain. Finally, for the flame index we see that although most of the reactions occur with negative flame index, there is also a significant amount of green zones, meaning positive flame index.
As reactivity is decreased, the amount of green zones are expected to increase until the green volume equals the purple volume.
For increasing reactivity, the opposite is expected. This can be seen by inspecting the lower row of \Fig{fig:snapshots_3D1_2600K}, which shows the flame index for case 3D1-2600K, which has significantly higher reactivity than case 3D1. For 3D1-2600K we do indeed see much less green coloring in the flame index, as expected. It is also interesting to see that there is less fine structure in this high-temperature flow. This comes from the fact that the temperature is higher, yielding a higher viscosity.

\begin{figure}[h!]
        \centering
        \includegraphics[width=1.0\textwidth]{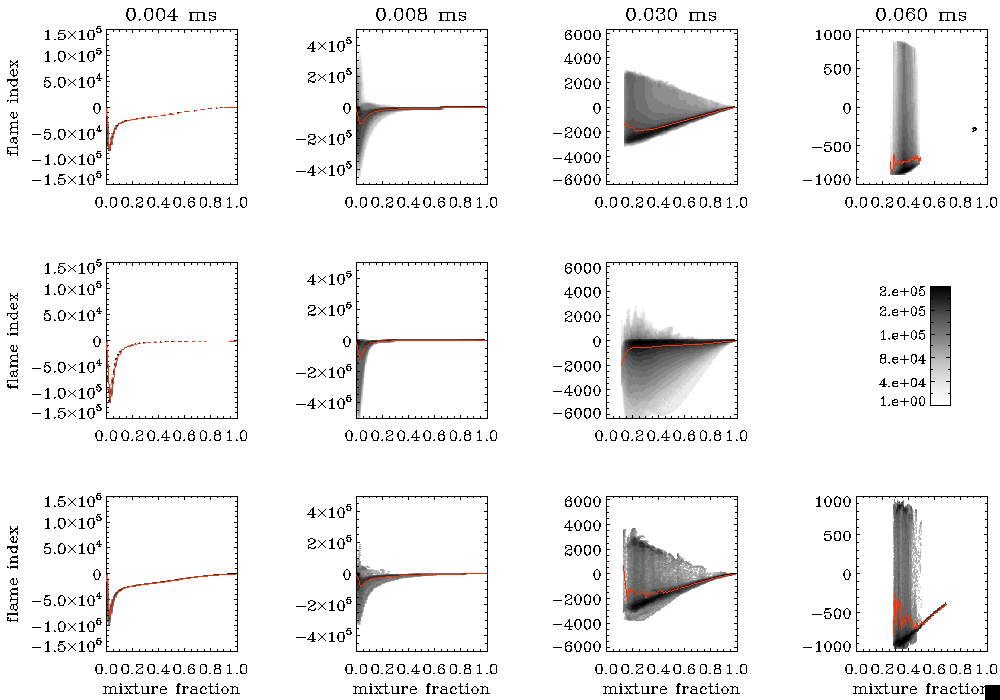}
        \caption{Flame index (FI) for 3D1 (upper row) 3D1-2600K (middle row) and 2D1 (lower row). The red lines represent the average values.}
        \label{fig:FI_dim}
\end{figure}

The upper row of \Fig{fig:FI_dim} shows a two-dimensional pdf of the flame index of simulation 3D1 as a function of mixture fraction for four different times. The legend of the pdf, corresponding to the number of cells with a given combination of FI and $z_H$, is show
at the far right of the second row in the figure.
At very early times, before any vortex generation in the boundary layer has occurred, we see from the left panel that the flame index is always negative. This is as expected since any reactions happening in the initial boundary layer between the two streams before mixing is initiated will proceed in non-premixed mode. As the time doubles to 0.008 ms, vortexes have already been generated from the shear and mixing has started. This is seen through the occurrence of the first positive contributions to the flame index. This only happens for very small mixture fractions, meaning that we are in very $\COO$-rich environments. The reaction-weighted average is nevertheless always negative, as represented by the red line. At $0.03\,$ms, mixing has become stronger and we now see an almost symmetric profile around zero, indicating that a significant fraction of the total volume is reacting in premixed mode. The reaction weighted average is nevertheless negative. It can also be observed that the probability to find grid cells in the volume with mixture fraction below 0.15 is zero. This is because mixing has forced hydrogen-containing species into grid cell in the domain. At the very latest time (the far-right panel) the pdf of the flame index is still rather symmetric, although it continues to be somewhat skewed towards negative values. At this point, mixing has narrowed the available mixture-fraction space to the range between 0.25 and 0.5.

The middle row of \Fig{fig:FI_dim} shows the PDF of the flame index as a function of mixture fraction for the case 3D1-2600K, which has a much larger Damk\"{o}hler number than the previous case (3D1). This results in the flame index profile being consistently dominated by negative values. Only at the latest time is there a hint of positive values, but these are just very weak. This supports the assumption that a highly reactive case as 3D1-2600K experience almost only reactions in non-premixed mode.

Finally we compare the flame index PDF of three and two-dimensional cases, by comparing the results for 3D1 (upper row) with that of 2D1 (lower row). From the figure it can be seen that the PDF's are remarkably similar, at least for the three first times. The main difference for these three times is that the three-dimensional results are significantly smoother due to better statistics (more mesh points). Also the results for the latest time resemble each other, with the clear exception for a surprising line extending into larger mixture fractions for the two-dimensional results. 
The reason for this discrepancy becomes apparent by comparing the contour plots of the mixture fraction and flame index for the two simulations, as shown in \Fig{fig:snapshots_6em5}. Here we see that the center of the large vortex of the two-dimensional simulation has higher mixture fraction than for the three-dimensional simulation. It can also be seen that the flame index inside this vortex is strongly correlated with the local mixture fraction. This results in the narrow peak in the pdf observed in the last panel of \Fig{fig:FI_dim}.

\subsubsection{The impact of initial $\OO$}
\begin{figure}[h!]
        \centering
        \includegraphics[width=1.0\textwidth]{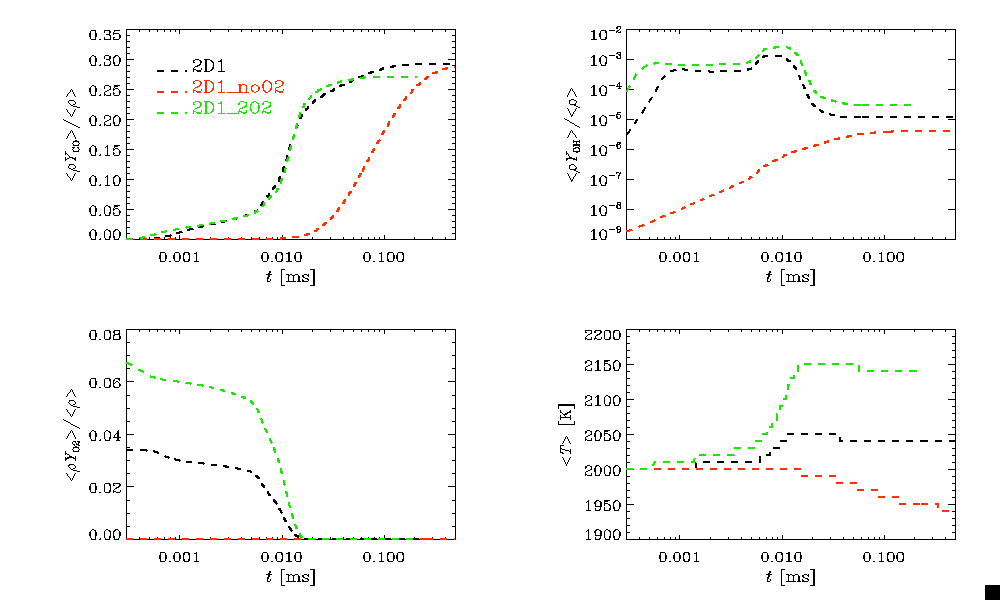}
        \caption{Results for simulations with various amounts of $\OO$ in the initial jet stream.}
        \label{fig:comp_runs_2000Kb_O2}
    \end{figure} 
It is interesting to see what happens when the $\OO$ in the jet is substituted with inert $\NN$, which is visualized in \Fig{fig:comp_runs_2000Kb_O2}. (Please note that the x-axis is logarithmic for this and the next figure.) From the figure it becomes clear that without $\OO$ in the jet, it takes 4-5 times longer to reach full yield of CO without any $\OO$ in the initial jet. In particular it can be seen that without $\OO$ in the mixture, it takes long to build a sufficiently large pool of OH, and it is not until $t\sim 0.015$ s that the OH pool is large enough to facilitate any significant production of $\CO$. Also, since the maximum level of OH is several orders of magnitude smaller without $\OO$ in the initial mixture, the maximum production rate of CO is also much lower (remember that the x-axis is logarithmic) than for the case with $\OO$. Given the significant benefit on conversion time of adding $\OO$ to the $\COO$ stream, it is tempting to increase the amount of $\OO$ even further. However, from the green line in \Fig{fig:comp_runs_2000Kb_O2}, it is clear that doubling the amount of $\OO$ does not result in any significant further speed-up compared to the original amount of $\OO$. This can be explained by the fact that, despite the amount of $\OH$ being somewhat higher for the case with double amount of $\OO$, the difference is not very large. Instead, we see that the final yield of $\CO$ is lower, since more of the $\HH$ is consumed by the additional $\OO$.
\begin{figure}[h!]
        \centering
        \includegraphics[width=1.0\textwidth]{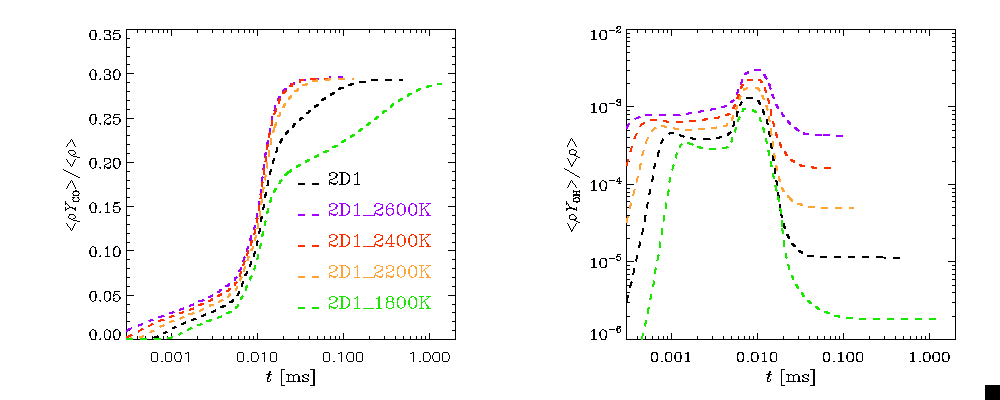}
        \caption{Results for simulations with various initial temperatures.}
        \label{fig:comp_runs_2000Kb_Temp}
\end{figure} 

\subsubsection{Temperature dependence}
The reactivity of the mixture is very sensitive to the initial temperature. In \Fig{fig:comp_runs_2000Kb_Temp}, the temporal evolution for simulations with different initial temperatures are shown. It is clear that reducing the temperature to 1800\,K, significantly increases the time to reach equilibrium (by around one order of magnitude). On the other hand, increasing the temperature to 2200\,K, reduces the time to reach equilibrium, but a further increase in temperature to 2400\,K, and eventually to 2600\,K, has little effect on the production rate of $\CO$. The reason that a further increase in temperature beyond 2200\,K does not result in a corresponding reduction in conversion time, is that for these high reactivities it is the turbulent mixing that limits the conversion, not the kinetic reactivity. This is also reflected in the Damk\"{o}hler number, which in \Tab{tab:params2} is shown to increase more than tenfold, from 0.0117 for the case with 2000\,K, to 0.198 for 2600\,K.

    \begin{figure}[h!]
        \centering
        \includegraphics[width=0.6\textwidth]{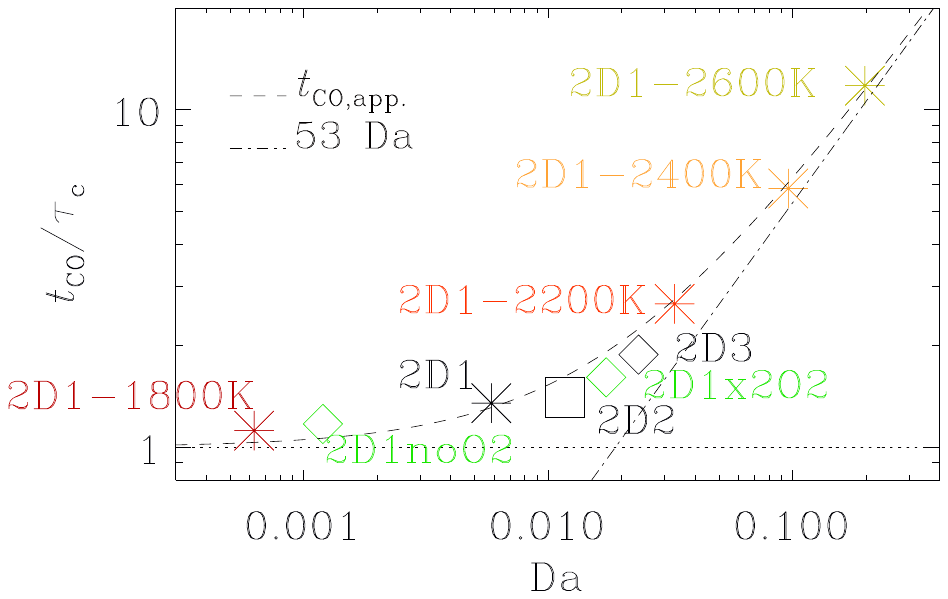}
        \caption{Normalized time to reach 95\% yield of CO as a function of Damk\"{o}hler number.}
        \label{fig:Da}
    \end{figure} 

Varying the relationship between the chemical ($\tau_c$) and fluid ($\tau_f$) timescales results in different Damk\"{o}hler numbers (see \Eq{eq:Da}). We know that for very large Damk\"{o}hler numbers, which means that reactions are mixing controlled, the time to reach equilibrium should scale linearly with the fluid timescale, $\tau_f$. For the opposite extreme, when the Damk\"{o}hler number is very small, the time to reach equilibrium should equal the chemical timescale, $\tau_c$. In the following, we approximate the time to reach equilibrium with the time it takes to reach a CO-yield of 95\% of the equilibrium yield, as represented by $t_{\rm CO}$. Based on the above, we end up with the following expressions for the value of $t_{\CO}/\tau_c$ for the two extremes of the Damk\"{o}hler number: 
\begin{equation}
\label{eq:tCO}
t_{\CO}/\tau_c=\left\{
\begin{array}{r@{\quad:\quad}l}
     1 & \Da \ll 1 \\
     \alpha\Da &  \Da \gg 1
\end{array}
\right.
\end{equation}
where $\alpha$ is a case dependent scaling factor. To test this expression we plot $t_{\CO}/\tau_c$ as a function of $\Da$ in \Fig{fig:Da}. In this figure, the horizontal dotted line corresponds to $t_{\CO}/\tau_c=1$, while the angle line dashed-dotted line represent $t_{\CO}/\tau_c=53\Da$, i.e.; $\alpha$ is 53. From the figure we clearly see that for the smallest and the largest Damk\"{o}hler numbers, the value of $t_{\CO}/\tau_c$ approach the limiting lines as defined in \Eq{eq:tCO}. For the current geometry, we therefore propose the following analytical expression, which has a maximum of 20\% error;
\begin{equation}
\label{eq:tCO_app}
    t_{\CO,{\rm app.}}=(1+\alpha \Da)\tau_c,
\end{equation}
which is visualized by the dashed line in \Fig{fig:Da}. The above expression does include all effects of variable Lewis numbers, which is critical when handling very light species such as hydrogen. Due to the weak endothermicity of the RWGS reaction, there are only small temperature gradients in these flows. The Soret effect is therefore not included.

\subsubsection{Reaction rates}
\begin{figure}[h!]
        \centering
        \includegraphics[width=1.0\textwidth]{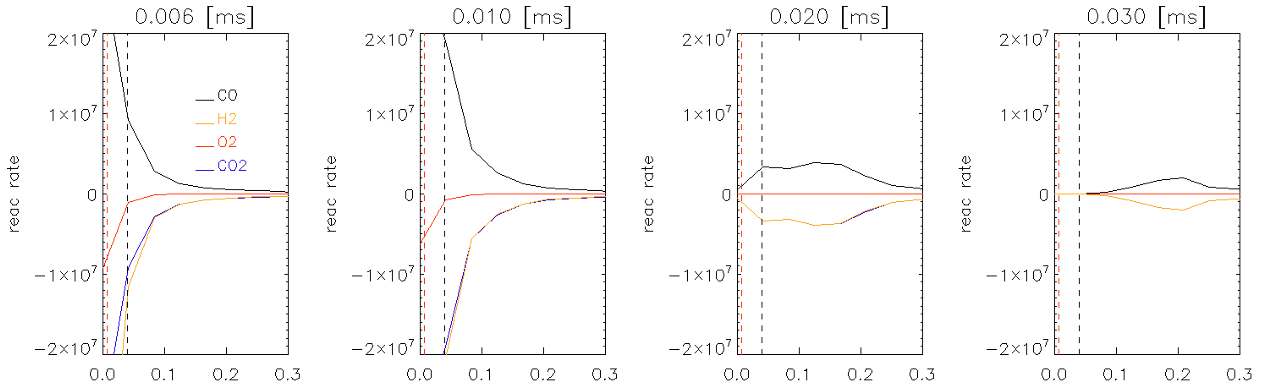}
        \caption{Reaction rates as a function of $z_H$ for simulation 2D1 as defined in \Tab{tab:params}. The vertical dashed black and red lines represent stoichiometric mixture fractions based on $\COO$ and $\OO$, respectively.}
        \label{fig:reac_rates_2D}
\end{figure}
It is known from non-premixed flames that the reaction rate typically peaks for mixture fractions around the stoichiometric mixture fraction. As we see from \Fig{fig:reac_rates_2D}, this is not the case for the RWGS reaction. Instead, we see a much broader distribution in reaction rates as function of mixture fractions, and the peak of these distributions is not related to the stoichiometric mixture fraction. This is yet another example where we see clear differences between the endothermic RWGS and flames.

\subsection{LES}
\label{sec:LES}

\begin{figure}[htbp]
    \centering
        \includegraphics[width=\textwidth]{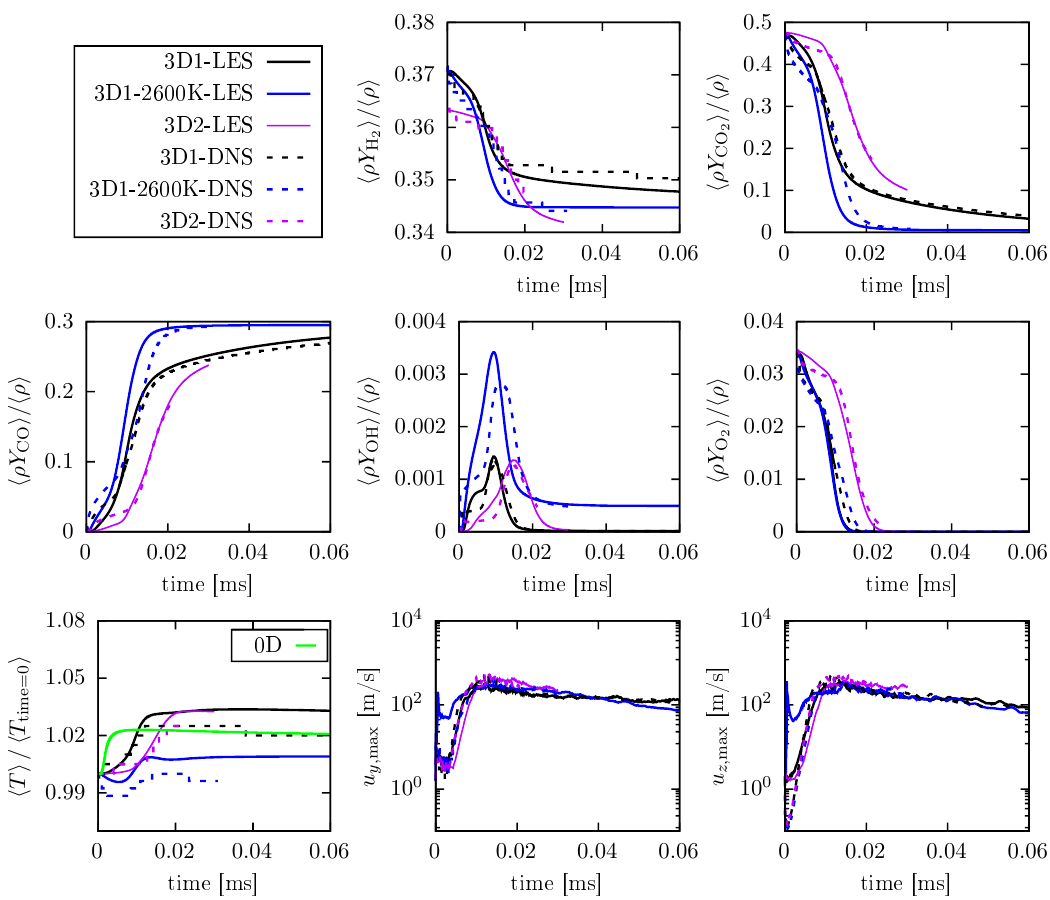}
    \caption{Integral quantities for the LES (solid lines) cases 3D1 (black), 3D1-2600K (pink) and 3D2 (blue) compared with the corresponding DNS results (dashed lines).}
    \label{fig:LES-results}
\end{figure}

In \Fig{fig:LES-results} we show the corresponding time traces computed by the LES as presented for the DNS in \Fig{fig:comp_runs_2000Kb}. Here, direct comparison of the three-dimensional cases between the LES (solid lines) and DNS (dashed lines) is offered. From the velocity time traces (lower panels) we observe excellent agreement on the shear-layer breakup time (about 0.01ms), a prerequisite for further comparison. It should be noted that the LES and DNS are initialized with similar noise levels ($~\sim 1\%$ of the jet-velocity) in all three directions, however, as seen from the maximum $z$-velocity component, the DNS notably features a strong decay of the artificial fluctuations, while the LES sustains the initial perturbations in the $z$ direction. Overall, the peak values and qualitative shape of the observed velocity traces agree reasonably well for the three cases considered.
Comparing the mass fraction integrals we see good agreement in the overall time evolution, specifically at later times, with the equilibrium correctly quantified in the LES. Further, the time to conversion, indicated by the $\CO$ integrals crossing the dotted 95$\%$ CO-yield line, is predicted to within 10$\%$ of the DNS results. During the initial phase (up to 0.01ms), the effect of choosing a finite $C_\mathrm{mix}$ in the LES model becomes apparent: In the DNS the entire computational cell is allowed to react, while in the LES the reaction rates are modulated, producing a slightly slower $\CO$ and $\OH$ production. The average temperature profiles predicted by the LES agree well with the DNS, since for all three cases, the difference relative to the initial value is $~\sim 1\%$.

\begin{figure}[htbp]
        \centering
        \includegraphics[width=0.6\textwidth]{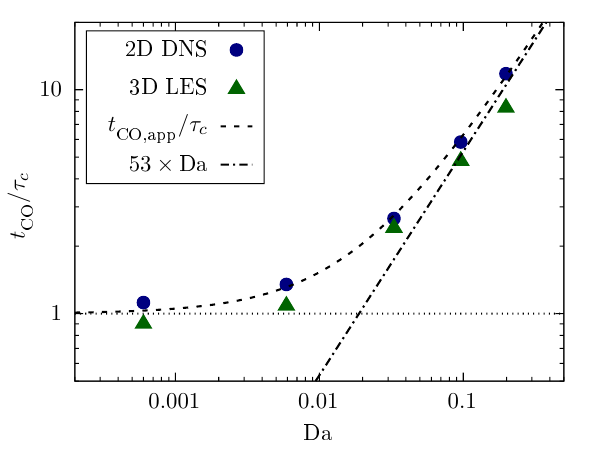}
        \caption{Normalized time to reach 95\% yield of CO as a function of Damk\"{o}hler number, $t_\mathrm{CO}/\tau_c$, as a function of the Damk\"{o}hler number Da, where $t_\text{CO,app}$ is from \Eq{eq:tCO_app}.}
        \label{fig:Da-LES}
\end{figure}

Figure \ref{fig:Da-LES} illustrates the normalized time required to reach 95\% CO yield, $t_\mathrm{CO}/\tau_c$, as a function of the Damk\"{o}hler number (Da), where $t_\text{CO,app}$ is from \Eq{eq:tCO_app}. The DNS and LES results show qualitative agreement, following the same trend with respect to Da. However, the LES results consistently lie slightly below the DNS results for all examined Da values. This discrepancy may be attributed to the higher average temperatures predicted by LES compared to DNS, as seen in \Fig{fig:LES-results}, since elevated temperatures accelerate chemical reactions and thus conversion. 
The same figure also includes results from a zero-dimensional (0D) simulation using the same chemical kinetics scheme, also performed in OpenFOAM, but without PaSR (green solid line in \Fig{fig:LES-results}). 
The initial condition of this 0D simulation is identical to the mass-averaged initial condition of the cases 3D1-DNS and 3D1-LES, and we see from \Fig{fig:LES-results} that the final temperature of the 0D case equals that of the 3D1-DNS case. Furthermore, the 0D simulation also yields a conversion time $t_\mathrm{CO}$ that is the same as obtained for 0D simulations with the Pencil-Code, supporting the interpretation that elevated temperatures in the three-dimensional LES are the reason for the lower $t_\mathrm{CO}$ found for 3D-LES simulations.

\begin{figure}[htbp]
    \centering
        \includegraphics[width=0.6\textwidth]{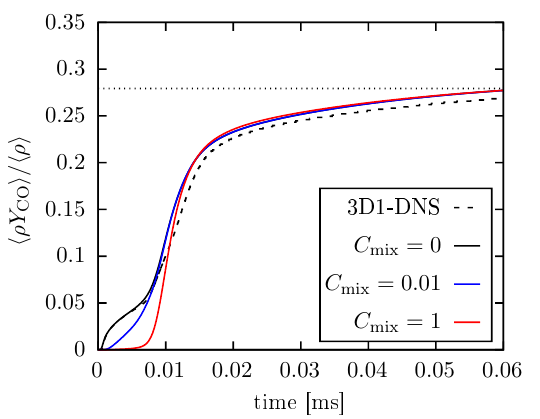}
        \caption{Evolution of the CO-yield for the case 3D1 and different values of $C_\mathrm{mix}$. }
    \label{fig:LES-3D1-Cmix}
\end{figure}

Figure \ref{fig:LES-3D1-Cmix} shows the CO production integral for three different values of $C_\mathrm{mix}$, compared to the DNS results. The values $C_\mathrm{mix} \in \{0,1\}$ correspond to $\tau_\mathrm{mix} \in \{0, \sqrt{\nu_\mathrm{eff}/\epsilon}\}$. Notably, when switching off the PaSR model ($C_\mathrm{mix} = 0$), and $\kappa = 1$, which considers the entire cell volume to be reactive, as in the DNS. The intermediate value ($C_\mathrm{mix} = 0.01$), represents the situation where cell resolution constraints are in place. 
We see that there are significant deviations during the initial phase, before the shear layer breaks up and turbulent mixing is negligible. After $\sim 0.012$ms, the integrals for different $C_\mathrm{mix}$ are indistinguishable, giving confidence that the chosen value of $C_\mathrm{mix} = 0.01$ can be considered reasonable to fix for all simulations. It should be noted that taking $C_\mathrm{mix} = 0$ gives the best agreement with the DNS for this specific study, however, as the goal is to test a \textit{viable} LES model for later use on an industrial scale reactor, the resolutions will vary, hence a value of $C_\mathrm{mix} = 0$ is unrealistic for future work; as $C_\mathrm{mix} = 0.01$ gives good agreement, we choose this value.

\begin{figure}[htbp]
    \centering
    \begin{subfigure}{0.16\textwidth}
        \includegraphics[width=\textwidth,keepaspectratio]{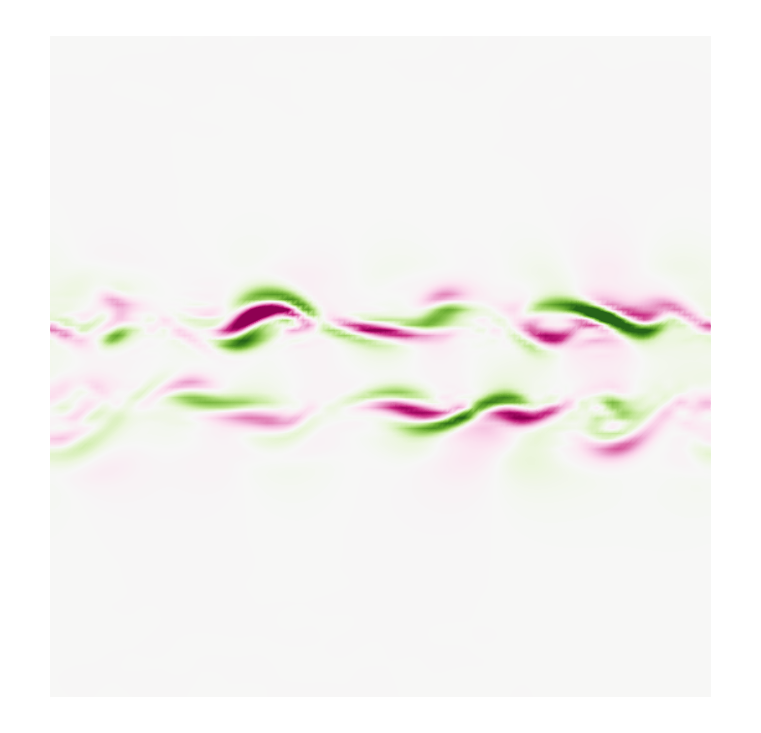}
    \end{subfigure}
    \hfill 
    \begin{subfigure}{0.16\textwidth}
        \includegraphics[width=\textwidth,keepaspectratio]{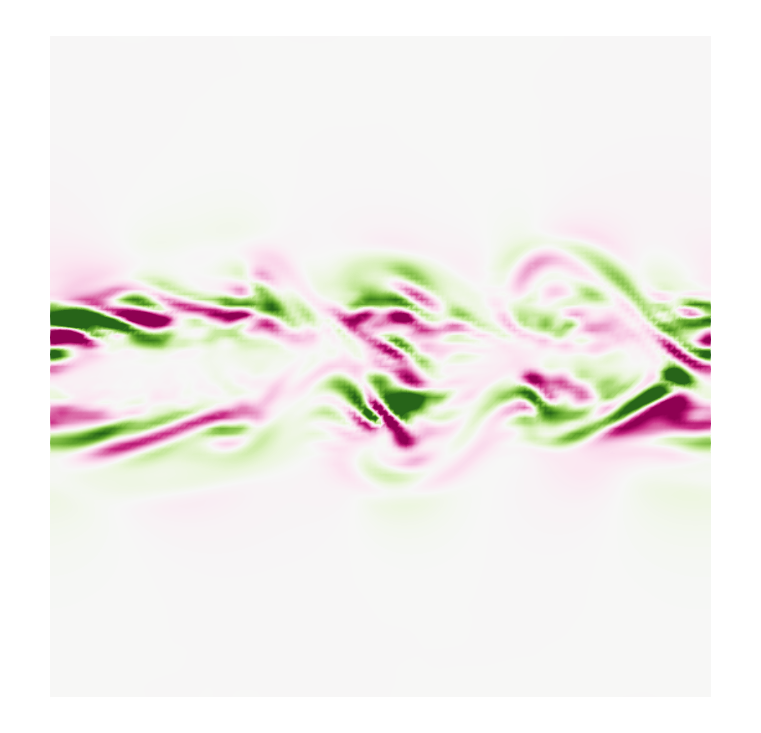}
    \end{subfigure}
    \hfill 
    \begin{subfigure}{0.16\textwidth}
        \includegraphics[width=\textwidth,keepaspectratio]{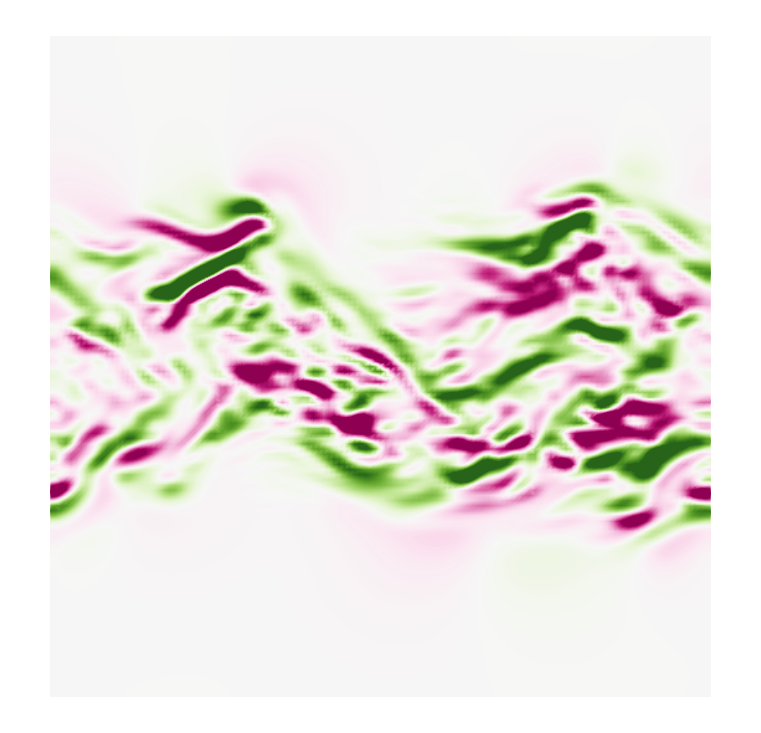}
    \end{subfigure}
    \hfill 
    \begin{subfigure}{0.16\textwidth}
        \includegraphics[width=\textwidth,keepaspectratio]{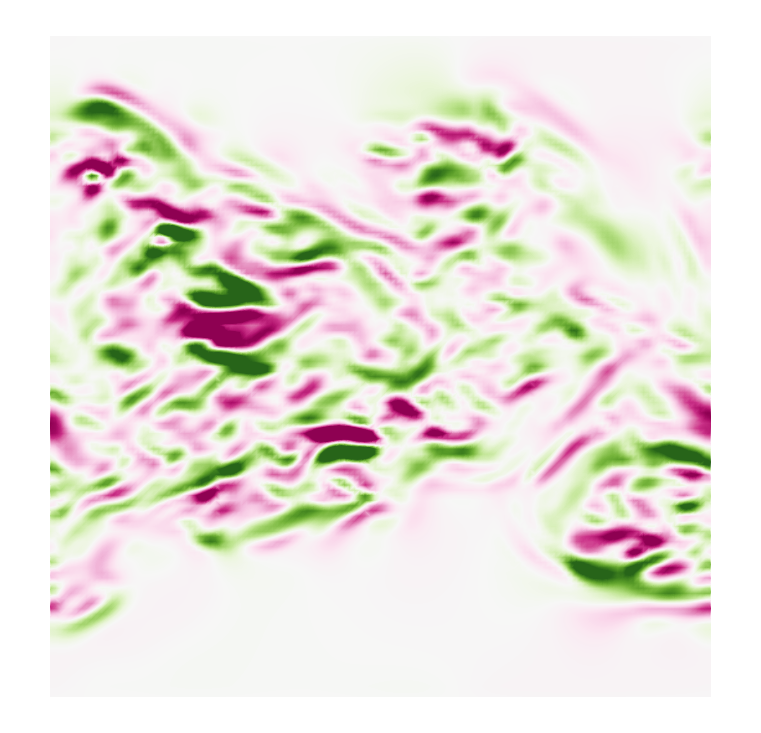}
    \end{subfigure}
    \hfill 
    \begin{subfigure}{0.16\textwidth}
        \includegraphics[width=\textwidth,keepaspectratio]{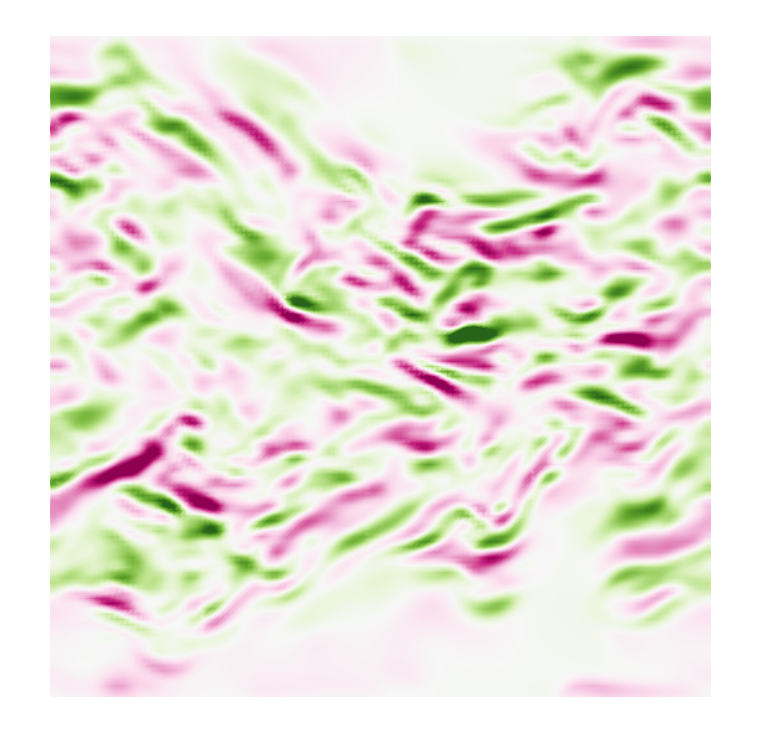}
    \end{subfigure}
    \hfill 
    \begin{subfigure}{0.16\textwidth}
        \includegraphics[width=\textwidth,keepaspectratio]{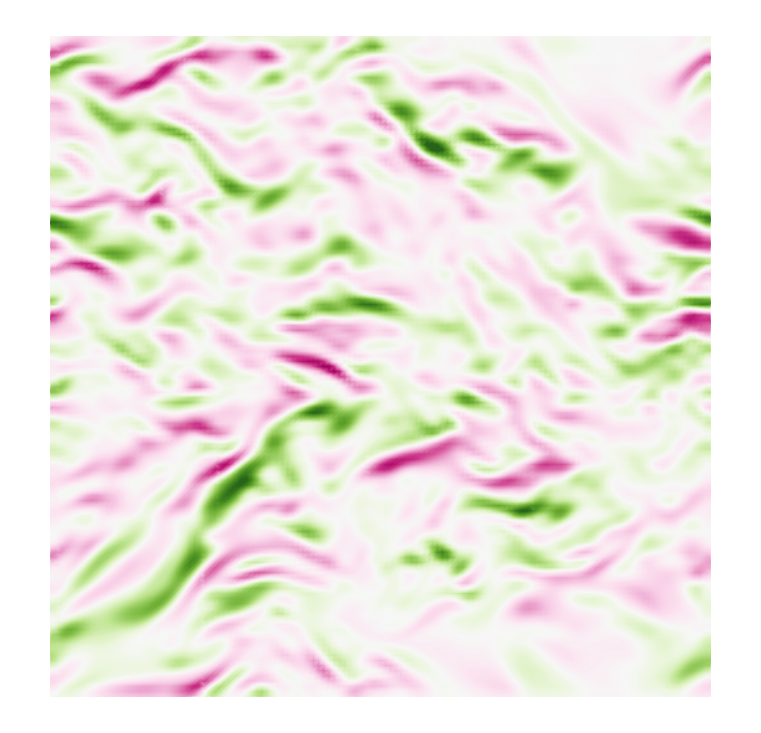}
    \end{subfigure}
    \\
    \begin{subfigure}{0.16\textwidth}
        \includegraphics[width=\textwidth,keepaspectratio]{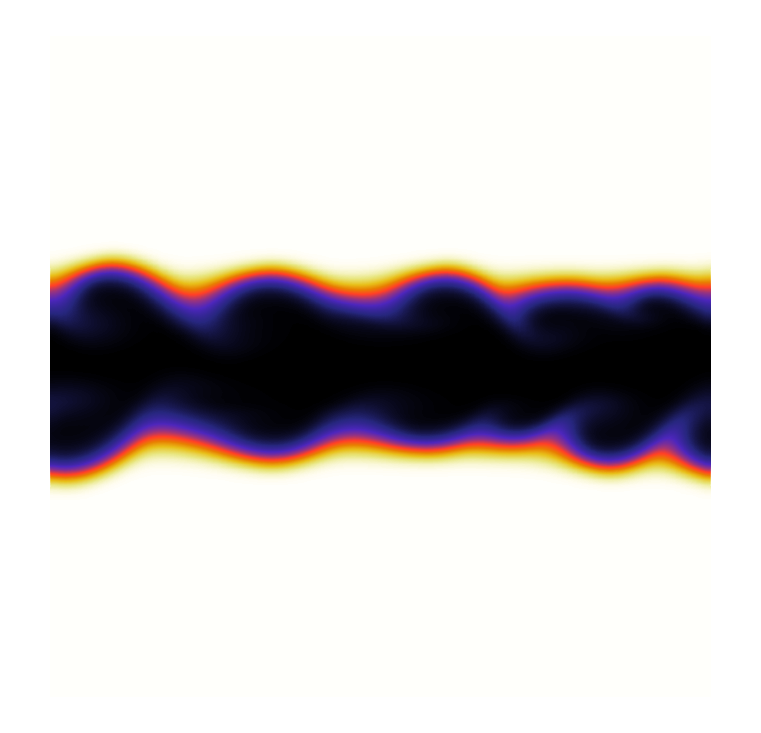}
    \end{subfigure}
    \hfill 
    \begin{subfigure}{0.16\textwidth}
        \includegraphics[width=\textwidth,keepaspectratio]{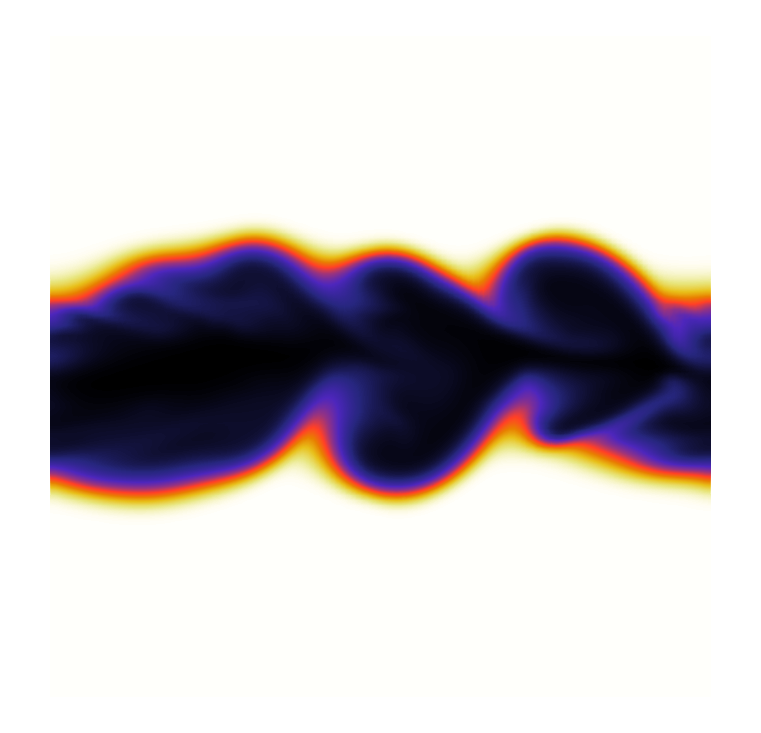}
    \end{subfigure}
    \hfill 
    \begin{subfigure}{0.16\textwidth}
        \includegraphics[width=\textwidth,keepaspectratio]{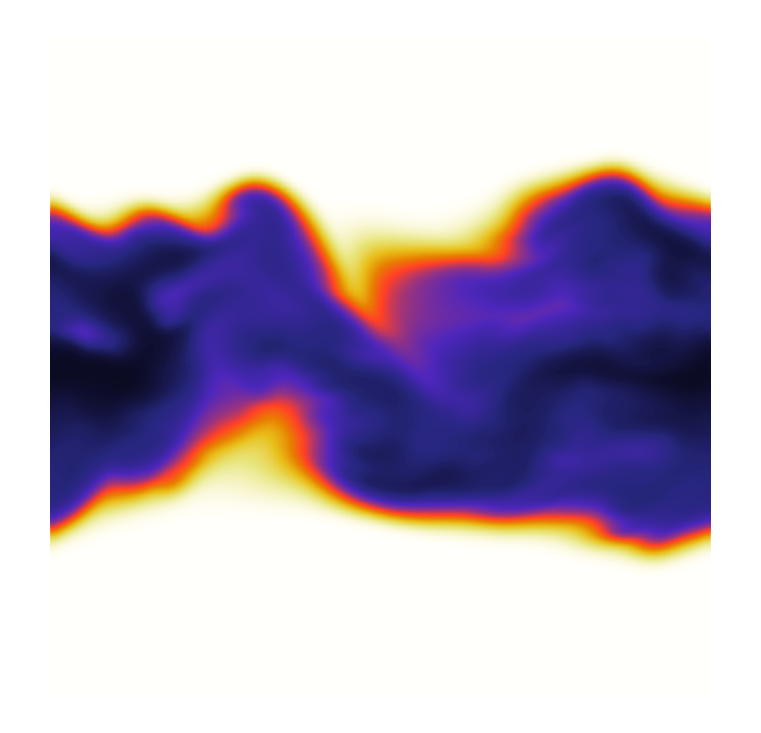}
    \end{subfigure}
    \hfill 
    \begin{subfigure}{0.16\textwidth}
        \includegraphics[width=\textwidth,keepaspectratio]{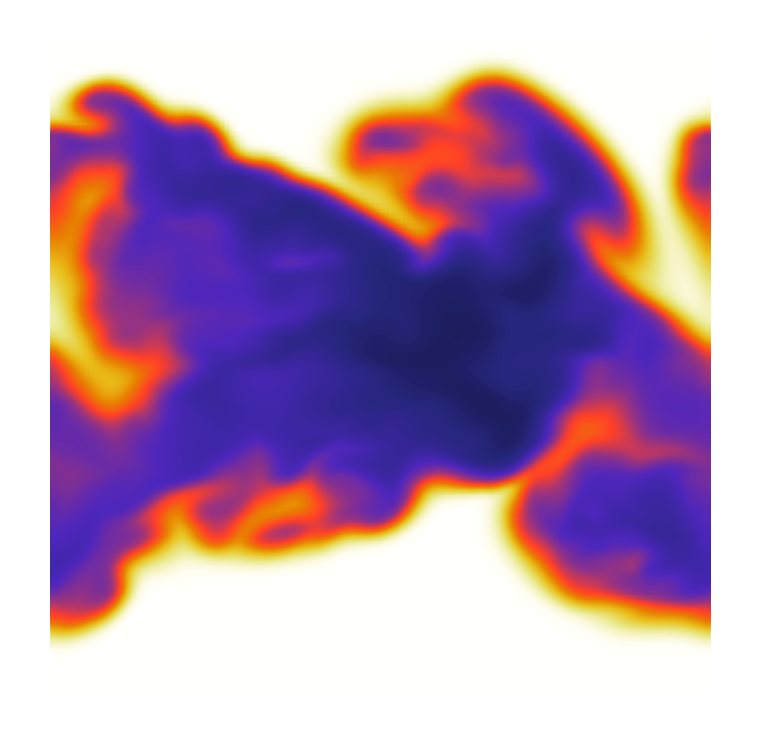}
    \end{subfigure}
    \hfill 
    \begin{subfigure}{0.16\textwidth}
        \includegraphics[width=\textwidth,keepaspectratio]{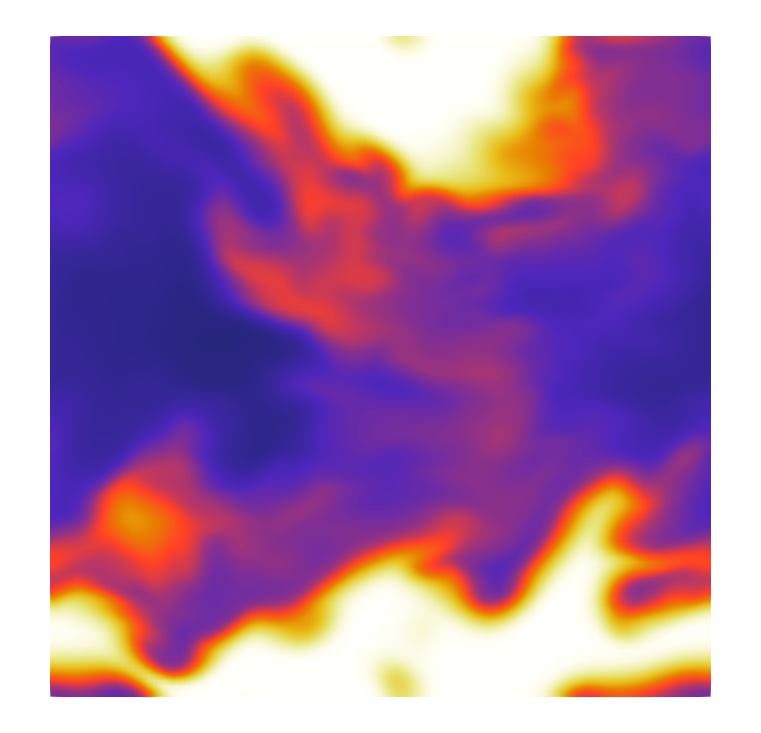}
    \end{subfigure}
    \hfill 
    \begin{subfigure}{0.16\textwidth}
        \includegraphics[width=\textwidth,keepaspectratio]{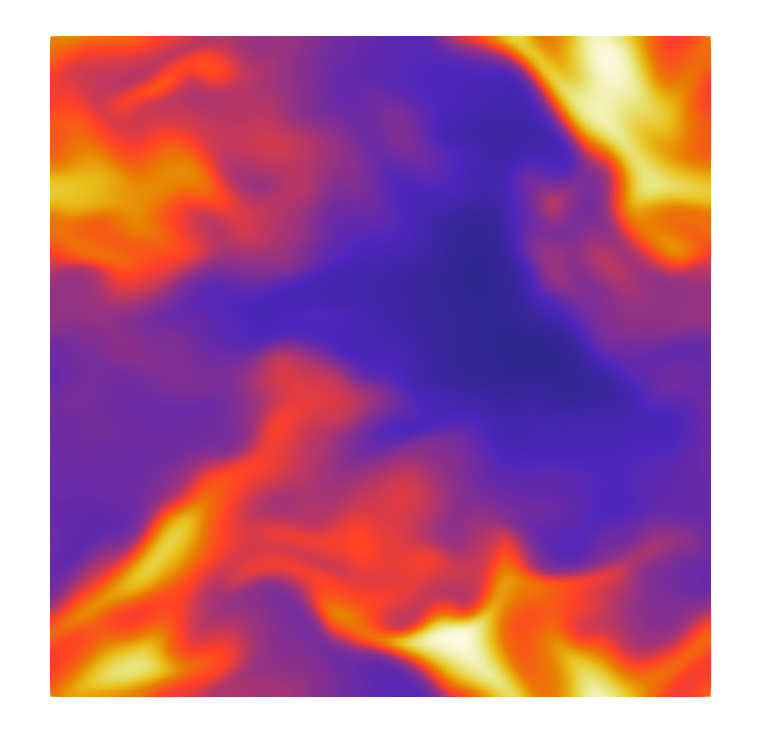}
    \end{subfigure}
    \caption{Evolution of the jet for the case 3D1 at
    $t=0.008$, 0.010, 0.015, 0.020, 0.025, and $0.030\,\textrm{ms}$
    illustrated by contour plots of the vorticity $x$-component (first row) and the mixture fraction (second row). }
    \label{fig:LES-3D1}
\end{figure}

\begin{figure}[htbp]
    \centering
    \begin{subfigure}{0.16\textwidth}
        \includegraphics[width=\textwidth,keepaspectratio]{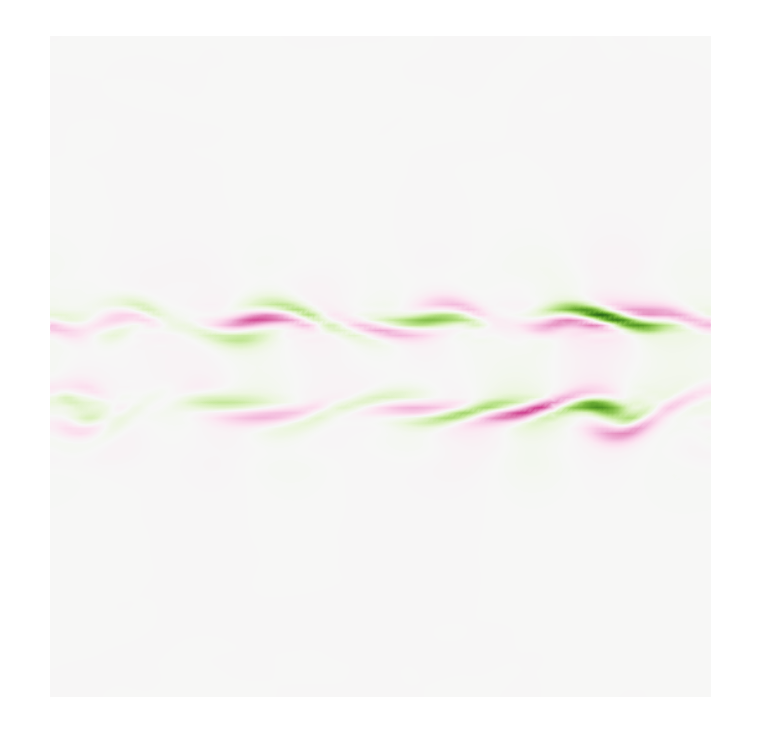}
    \end{subfigure}
    \hfill 
    \begin{subfigure}{0.16\textwidth}
        \includegraphics[width=\textwidth,keepaspectratio]{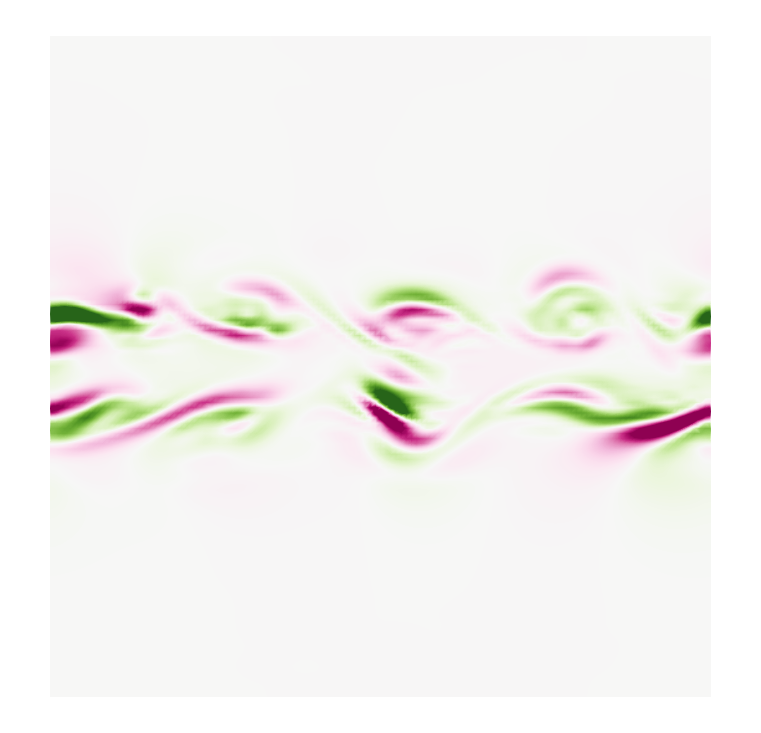}
    \end{subfigure}
    \hfill 
    \begin{subfigure}{0.16\textwidth}
        \includegraphics[width=\textwidth,keepaspectratio]{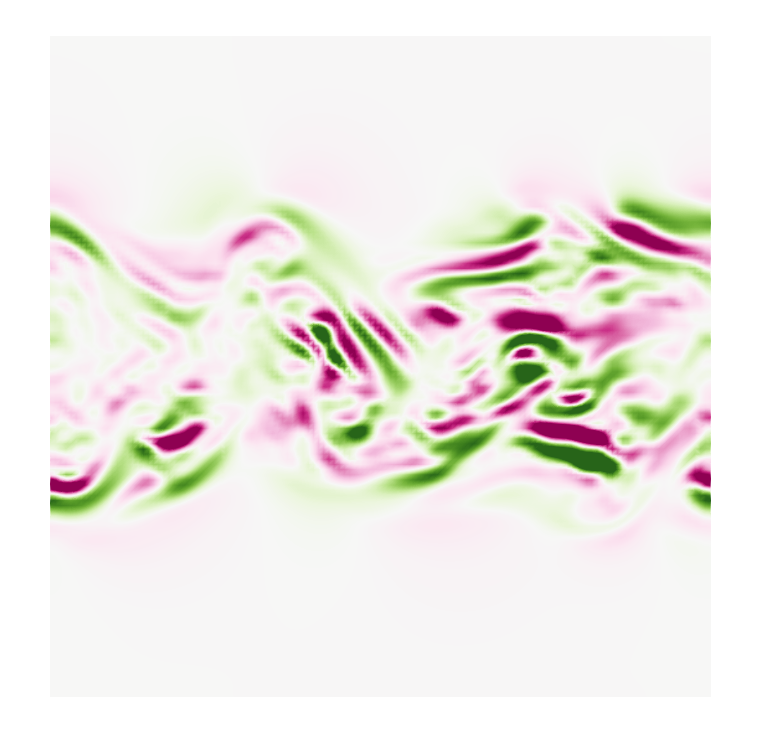}
    \end{subfigure}
    \hfill 
    \begin{subfigure}{0.16\textwidth}
        \includegraphics[width=\textwidth,keepaspectratio]{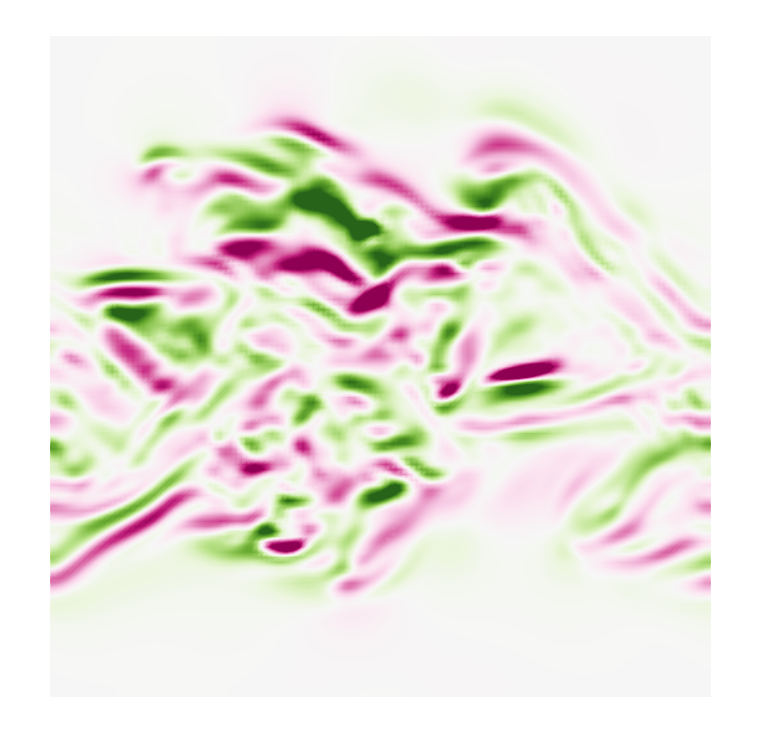}
    \end{subfigure}
    \hfill 
    \begin{subfigure}{0.16\textwidth}
        \includegraphics[width=\textwidth,keepaspectratio]{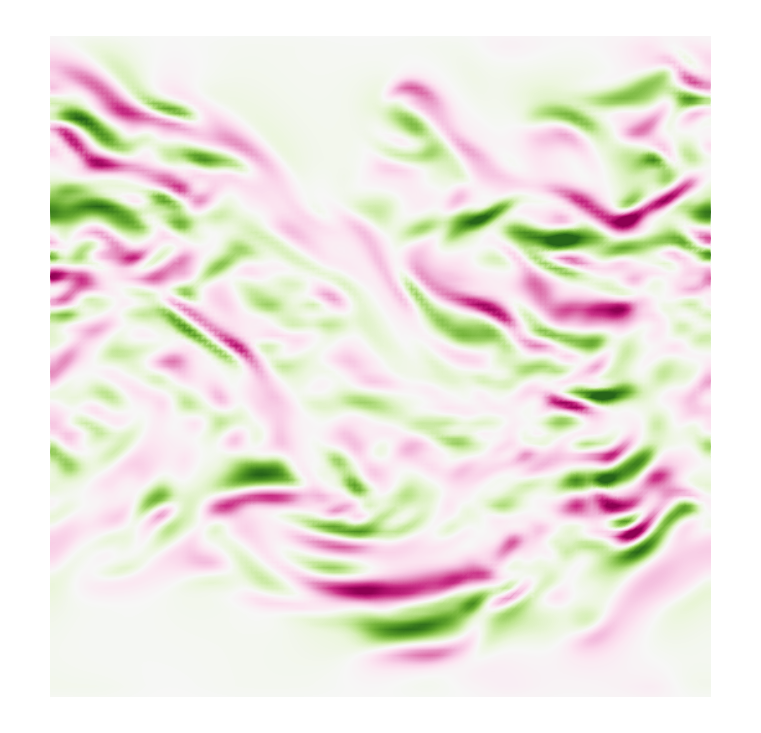}
    \end{subfigure}
    \hfill 
    \begin{subfigure}{0.16\textwidth}
        \includegraphics[width=\textwidth,keepaspectratio]{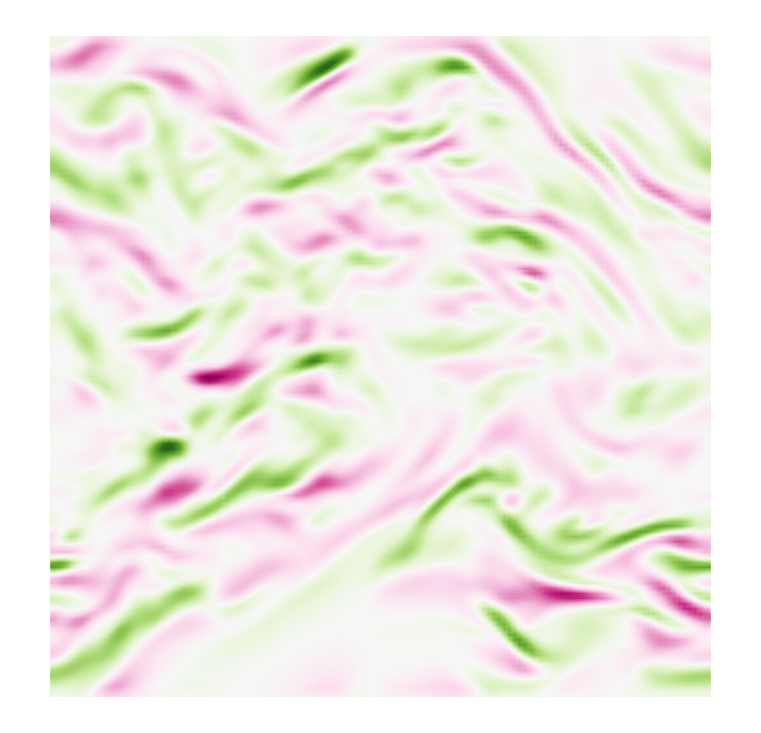}
    \end{subfigure}
    \\
    \begin{subfigure}{0.16\textwidth}
        \includegraphics[width=\textwidth,keepaspectratio]{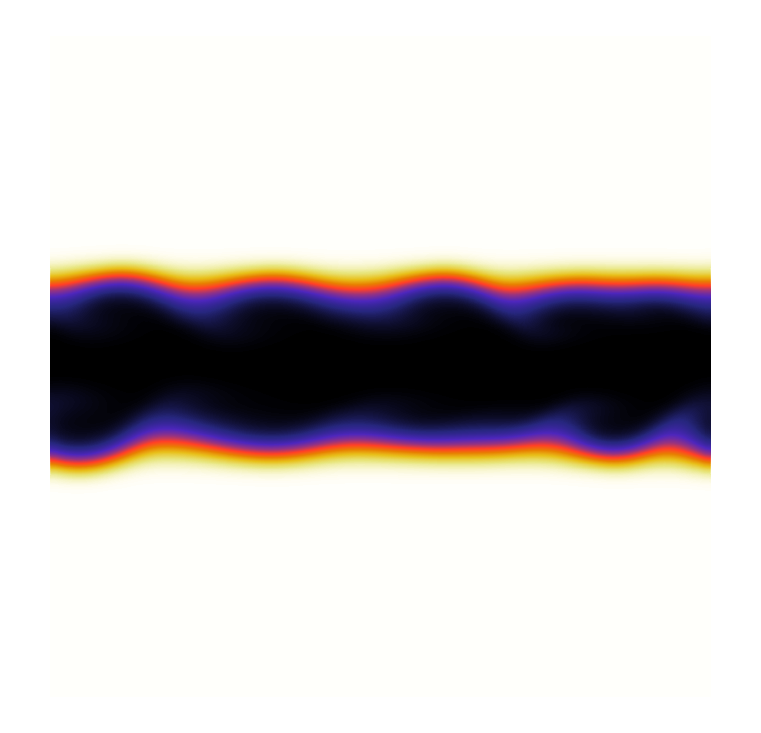}
    \end{subfigure}
    \hfill 
    \begin{subfigure}{0.16\textwidth}
        \includegraphics[width=\textwidth,keepaspectratio]{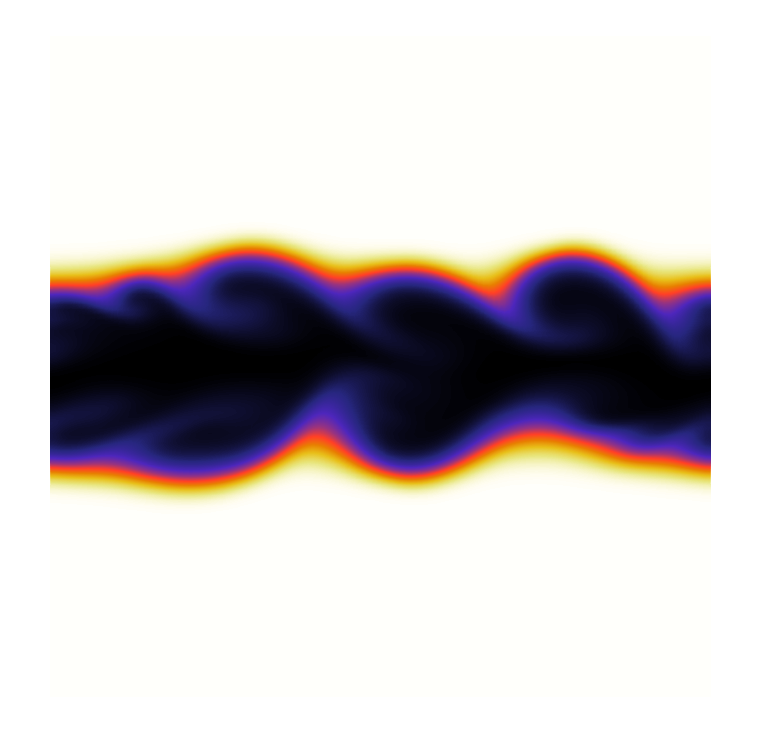}
    \end{subfigure}
    \hfill 
    \begin{subfigure}{0.16\textwidth}
        \includegraphics[width=\textwidth,keepaspectratio]{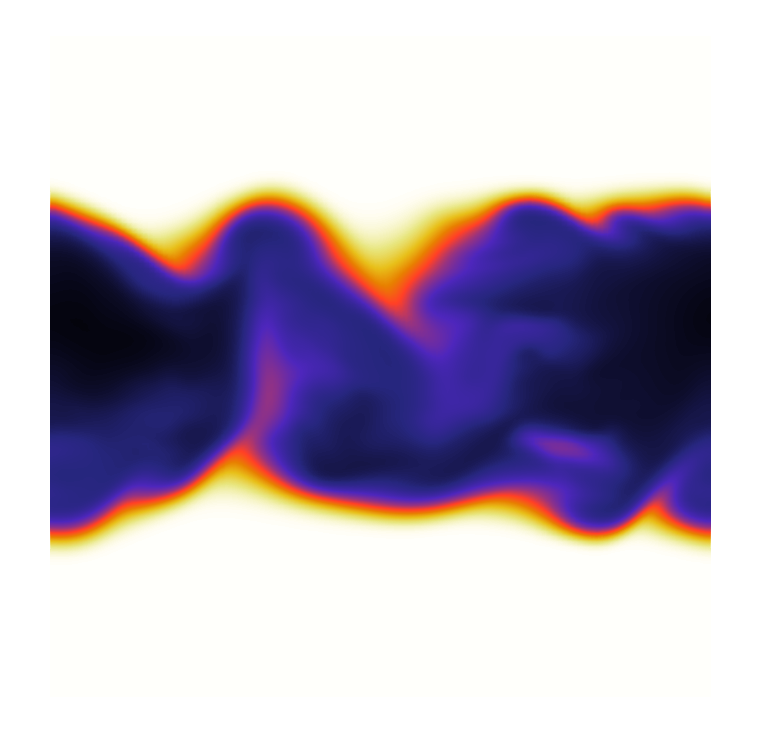}
    \end{subfigure}
    \hfill 
    \begin{subfigure}{0.16\textwidth}
        \includegraphics[width=\textwidth,keepaspectratio]{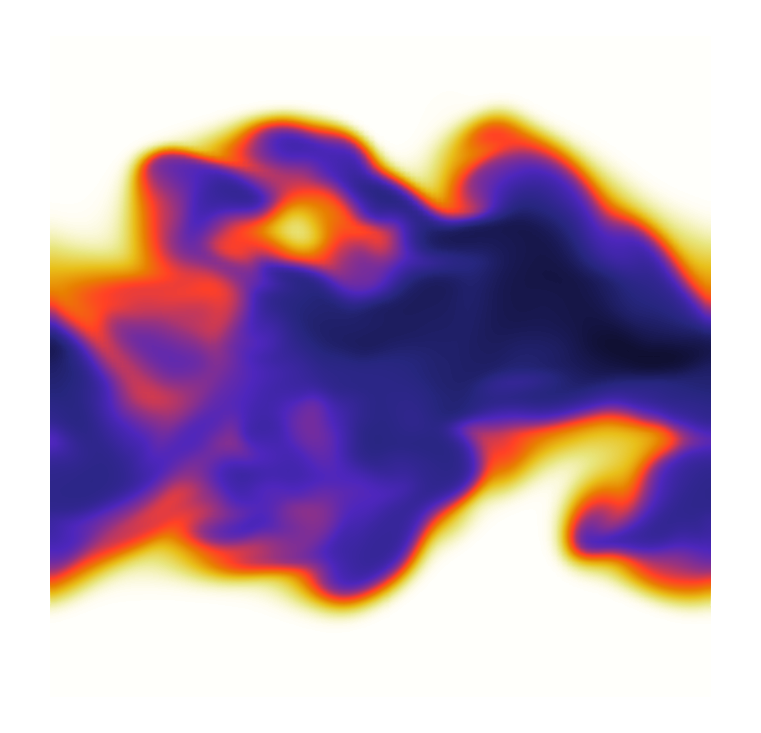}
    \end{subfigure}
    \hfill 
    \begin{subfigure}{0.16\textwidth}
        \includegraphics[width=\textwidth,keepaspectratio]{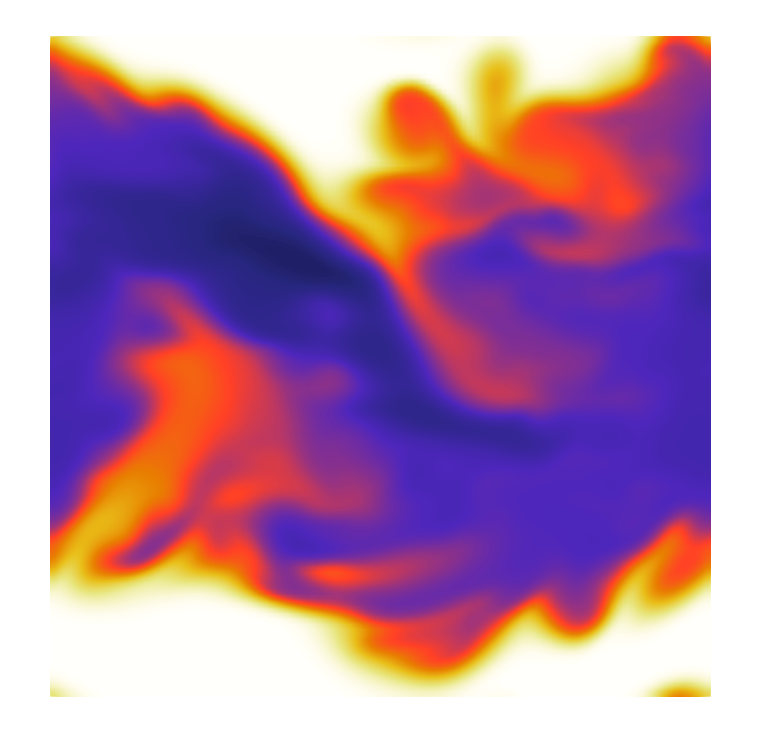}
    \end{subfigure}
    \hfill 
    \begin{subfigure}{0.16\textwidth}
        \includegraphics[width=\textwidth,keepaspectratio]{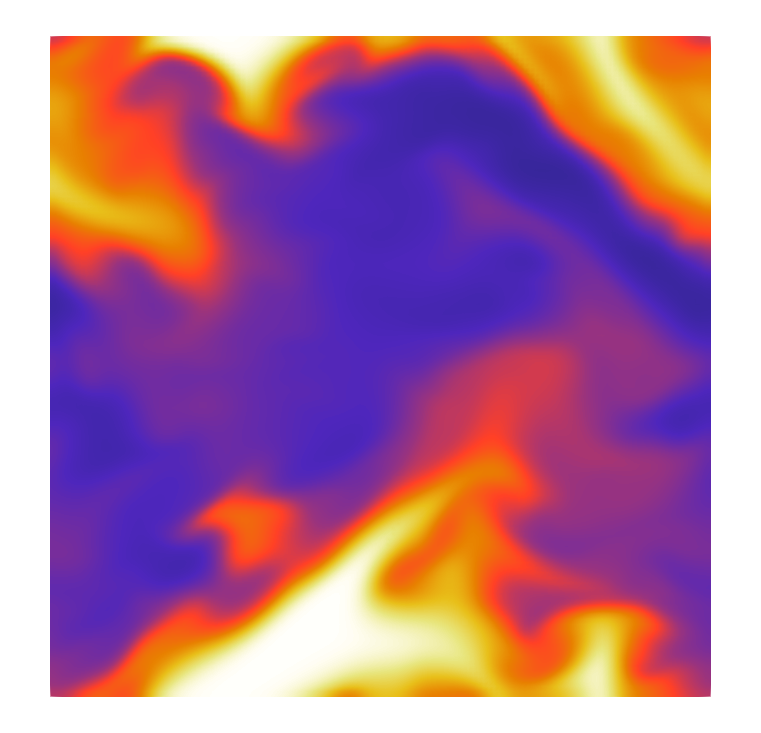}
    \end{subfigure}
    \caption{Evolution of the jet for the case 3D1-2600K at
    $t=0.008$, 0.010, 0.015, 0.020, 0.025, and $0.030\,\textrm{ms}$
    illustrated by contour plots of the vorticity $x$-component (first row) and the mixture fraction (second row).}
    \label{fig:LES-3D1-2600K}
\end{figure}

Contour plots of the vorticity $x$-component and the mixture fraction as computed by the LES are illustrated in \Fig{fig:LES-3D1}. There is overall qualitative agreement in the time evolution of the flow field in terms of jet-breakup dynamics and $\HH$ conversion, cf. \Fig{fig:snapshots_3D1}, showing the mixture fraction with black color denoting zero mixture fraction (no elemental hydrogen) and white representing a mixture fraction of unity (co-flow composition).
Similarly to the evolution of the DNS jet, we see that after approximately 0.03\,ms there is almost no pure $\HH$ left in the domain.
From the mixture fraction snapshots we also observe the same trend in terms of slightly slower dynamics for the 2600\,K vs 2000\,K simulations, as in the DNS. We also clearly observe the filtering effect of the LES, yielding less fine structures than the DNS. Similarly to the DNS results, the higher viscosity in the 3D1-2600K case, shown in \Fig{fig:LES-3D1-2600K}, also produces larger flow structures.

\newpage

\section{Conclusions}

Comparing the reaction rates of various syngas mechanisms under conditions relevant to the RWGS reaction
shows that all the detailed syngas mechanisms studied here yield comparable results.
Global reaction mechanisms should not be used though, since they do not account for the forward water-gas-shift reaction, and hence, they are not able to predict the equilibrium. For this study the two smallest mechanisms are therefore used, since they are more computationally efficient than the larger ones.
At atmospheric pressure, it is found that just very minor traces of $\OO$ can have a strong effect on the rate of the RWGS reaction. Since existing experimental tests of the non-catalytic reverse water gas shift 
reaction do not provide data on the level of $\OO$ at the inlet at sufficiently high accuracy, it is not possible to validate the performance of the syngas reaction mechanisms against the currently available experimental data. At high pressures and temperatures, relevant for non-catalytic RWGS, the free-radical promoted effect of $\OO$ is, however, less pronounced than at atmospheric pressure.

The temporal jet framework is used to investigate the turbulence-chemistry interactions of the RWGS reaction.
Form the flame index it is found that for an initial temperature of 2600\,K,
essentially all reactions proceed in non-premixed mode, while already at a temperature of 2000\,K,
a significant fraction of the $\COO$ conversion proceeds in a premixed mode. 

At 30 bars, having 5\% $\OO$ in the $\COO$-stream results in a 2--3 times shorter conversion time than without $\OO$.
Doubling the amount of $\OO$ beyond this has, however, little effect on the
conversion rate except that less $\CO$ is produced due to the extra oxidation.

For the temporal jet configuration studied in this paper, an algebraic description
of the time to reach 95\% conversion of $\CO$ is proposed; see \Eq{eq:tCO_app}.
The resulting expression for $t_{\CO}$, which is a function of the Damk\"{o}hler number and the chemical timescale,
accounts for the effects of Reynolds number, oxygen level, and temperature (as can be seen from \Fig{fig:Da}).
As such, this equation can be used to \textit{a priori} give an estimate of the conversion time for turbulent shear flows.

Finally, we compared the DNS results with those of the corresponding LES,
which use a dynamic one-equation subgrid-scale eddy viscosity model,
together with the PaSR model for turbulence chemistry interactions.
Despite the fact that the PaSR model is developed with strongly exothermic flames in mind,
the comparison shows that the current LES-setup is able to nicely reproduce the DNS results even for the endothermic RWGS reaction.

The present paper shows that, by comparing with DNS, the LES with the PaSR model can be used for endothermic reactions at small scales.
Recommended additional work is then to validate the combination of turbulence modeling and chemical kinetics used in LES against experimental results at larger scale. For proper validation it is critical that all boundary conditions of the experiments are well known and that sufficient measurements data are obtained.

\section*{Acknowledgments}
Funding from the Norwegian research council through the SAFIRE project (346574) is acknowledged.

\bibliography{library}
		
\end{document}